\documentclass[a4paper,fleqn,usenatbib]{mnras}
\usepackage[T1]{fontenc}
\usepackage{newtxtext,newtxmath}
\usepackage{ae,aecompl}
\usepackage{graphicx}	
\usepackage{amsmath}	
\usepackage{hyperref}
\usepackage{microtype}
\usepackage{verbatim}
\usepackage{graphicx}
\usepackage{subfig}
\usepackage{multirow}
\usepackage{array}
\usepackage{url}

\DeclareRobustCommand{\VAN}[3]{#2}
\let\VANthebibliography\thebibliography
\def\thebibliography{\DeclareRobustCommand{\VAN}[3]{##3}\VANthebibliography}

\title[Longslit spectroscopy of NGC 1313 X-2]{Deep optical spectroscopic monitoring of the pulsating ULX NGC 1313 X-2 with longslit Gemini observations}

\author[Rajath Sathyaprakash et al.]{Rajath Sathyaprakash$^{1}$$^{2}$\thanks{E-mail: rajath.sathyaprakash@gmail.com} and Timothy. P. Roberts$^{2}$ 
\\
$^{1}$Scuola Universitaria Superiore IUSS Pavia, Palazzo del Broletto, piazza della Vittoria 15, I-27100 Pavia, Italy\\
$^{2}$Centre for Extragalactic Astronomy, Durham University, Dept of Physics, South Road, Durham DH1 3LE, UK\\
}

\date{Accepted XXX. Received YYY; in original form ZZZ}

\pubyear{2026}

\begin{document}
\label{firstpage}
\pagerange{\pageref{firstpage}--\pageref{lastpage}}
\maketitle

\begin{abstract}
This study reports on archival long--slit {\it{Gemini-South}} spectroscopic observations of the pulsating ULX NGC 1313 X--$2$ in order to classify the stellar type of the donor star in the binary, study the diffuse nebula and other emission components. We stack flux--calibrated spectra from twelve nights of observations, finding that it features a marginal Balmer break below 3800 \AA. We then fit the stacked spectrum with model grids from standard stellar atlases to find that the continuum is best characterised by an evolved A--type supergiant. Using the inferred stellar radius and metal abundance, we update on constraints of the orbital parameters of the binary system. We also compute the lag--frequency spectrum to find a $2 \pm 0.1$ day--timescale lag between the X--ray and optical continuum over the period spanning MJD 59125--MJD 59195, suggesting that the optically emitting region and the ULX environment is more complex than a standard thin disc. We corroborate on results from earlier studies for the gas bubble expansion rates by modelling the [O III $\lambda 4959, 5007$] emission line profiles to constrain the kinetic power of radiatively driven gas outflows in comparison to the accretion power. In summary, we expand on previous efforts to investigate the formation history of the binary using multi--wavelength observations.
\end{abstract}


\section{Introduction}

Ultraluminous X--ray sources are hypothesised to belong to an extreme class of young high mass X--ray binary systems, as empirically defined by their apparent luminosities to be in excess of the Eddington limit for a 10$M_{\odot}$ black hole (for a recent review see \citealt{kingmiddleton23}; \citealt{pintowalton23}; \citealt{pintokosec23}). Recent population studies (\citealt{earnshaw2019}; \citealt{kovlakas2020}; \citealt{walton2022}), aided mainly by pointed X--ray observations, but also optical and radio instruments, have highlighted that many ULXs tend to reside in regions of distant spiral galaxies associated with recent star formation and young stellar populations, as implied by the discovery of several ULXs in the Antennae (\citealt{fabbiano2001}) and in NGC 4485--90 (\citealt{roberts2002} and \citealt{gladstone2009}). But, some ULXs are associated with early type low metallicity systems such as dwarf galaxies (\citealt{swartz2008}; \citealt{walton2011}; \citealt{brorby2014}) and globular clusters, providing further credence to the scenario that they represent the high luminosity distribution of low mass X--ray binary systems (\citealt{soria2005}; \citealt{bregman2006} and \citealt{thygesen2023}). If one assumes isolated binary evolution, the preference of ULX locations toward lower metallicity environments could suggest a correlation with the mass of the accretor, which naturally suggests higher luminosities, implies supression of stellar winds in the late evolutionary phases of the companion star, leading to smaller orbits and a higher probability to form ULXs (\citealt{kovlakas2020}). 

\begin{figure*}
\includegraphics[scale=0.5]{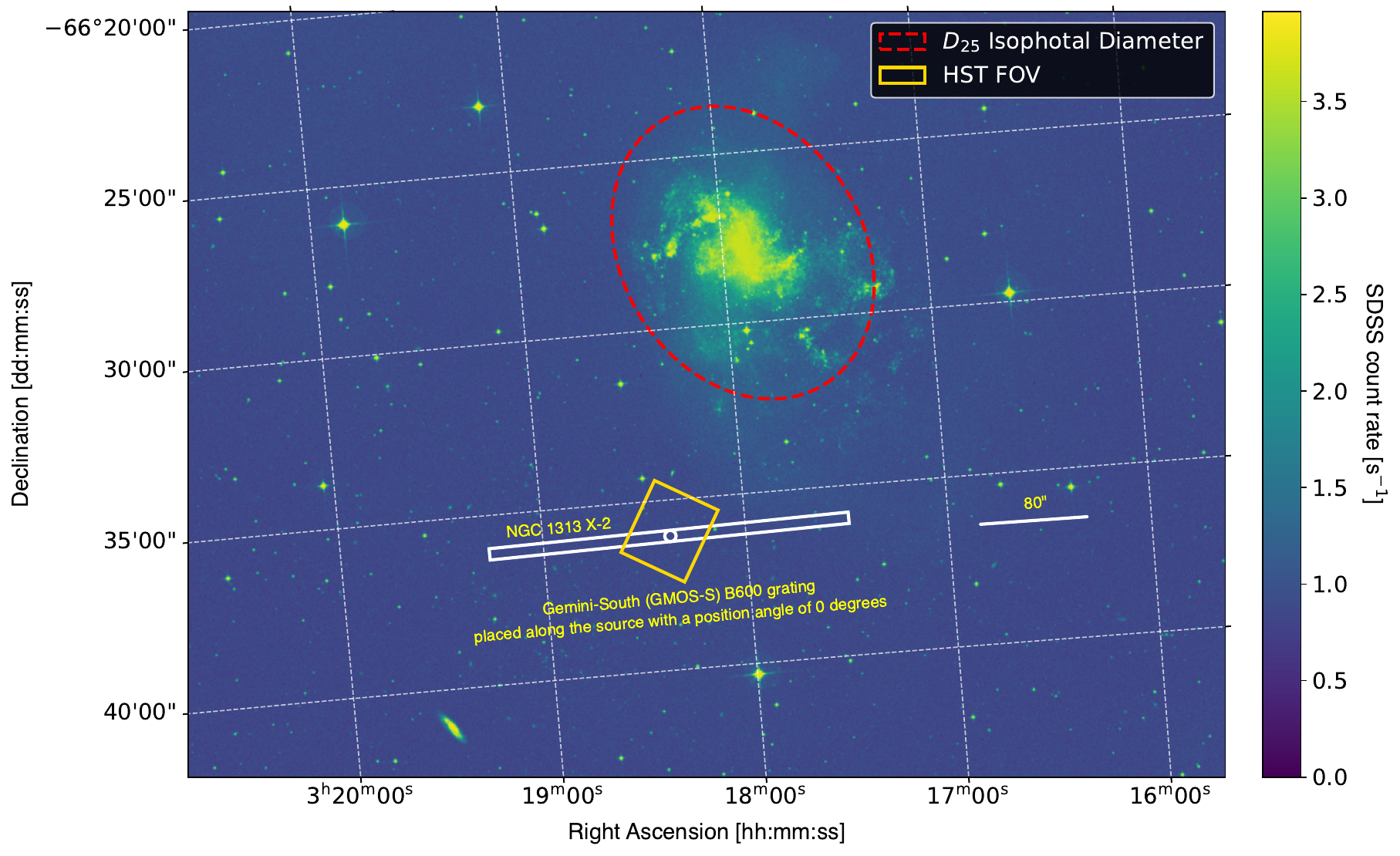}
	\caption{We present the Sloan Digital Sky Survey (SDSS) image of the galaxy NGC 1313, known to host two ultraluminous X--ray sources: X--1 and X--2. Here, we study the pulsating ULX NGC 1313 X--2, found towards the southern outskirts of the galaxy, outside the D25 isophotal ellipse. The sky--location of the source $\alpha = 03^{\mathrm{h}}\,18^{\mathrm{m}}\,22.3^{\mathrm{s}}$ and $\delta = -12^{\circ}\,34^{\prime}\,56.7^{\prime\prime}$ is marked by the solid white circle with radius of 5 arc--seconds. The configuration of the {\it{Gemini}--South} B600 grating and 2015--2016 {\it{HST}}--footprint (orange box) is also overlaid. The colour--bar shows the SDSS count-rate in units of s$^{-1}$.}
\label{fig:galaxy_image_ngc1313}
\end{figure*}

\begin{figure*}
\includegraphics[scale=0.4]{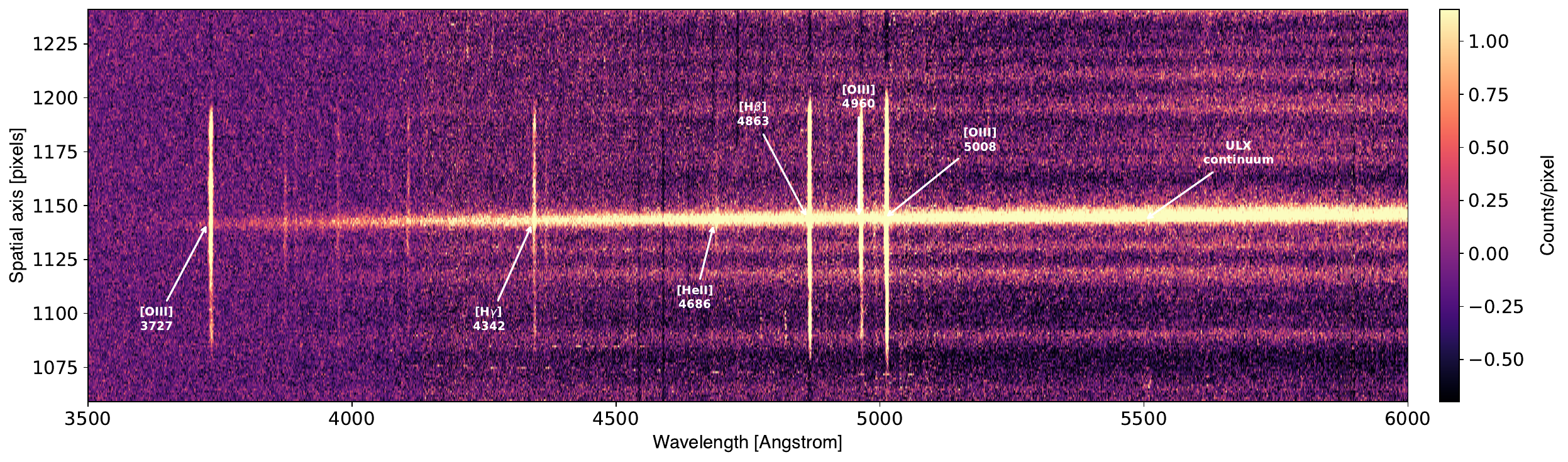}
\caption{The 2D stacked long--slit GMOS-S optical spectrum of NGC 1313 X-2, with the colour--bar showing the instrumental counts per pixel, combining observations from all ten nights in 2009. The spectrum features several emission lines, a majority of which are associated with the bubble nebula surrounding the ULX, while the highly variable [He II] $\lambda$4686 line is presumed to originate from the ULX counterpart (\citealt{roberts2011}).}
\label{fig:stacked_spec2D}
\end{figure*}

\begin{figure*}
\includegraphics[scale=0.5]{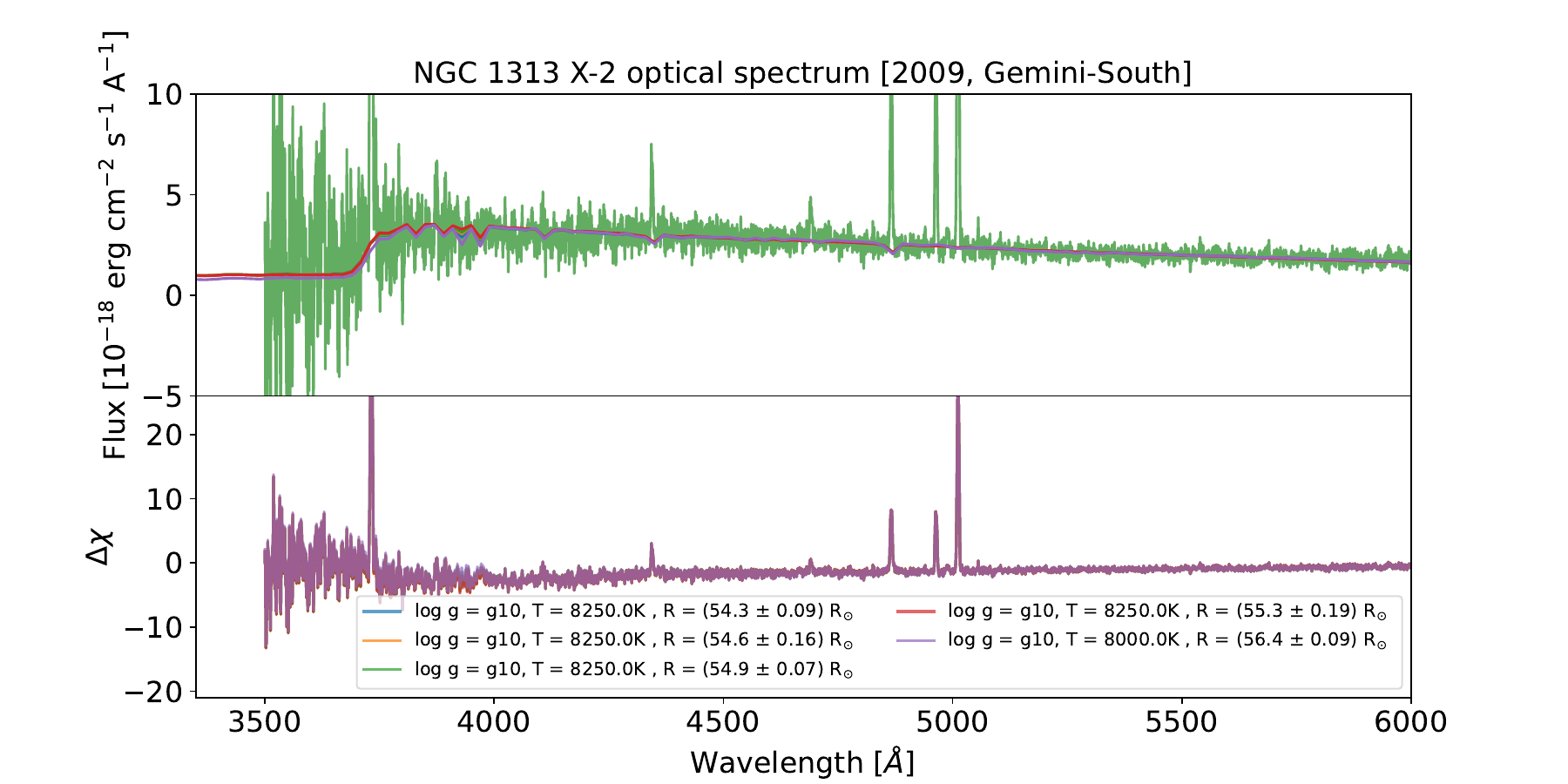}
\caption{{\it{Top panel}:} The flux calibrated 1D stacked GMOS--S optical spectrum of NGC 1313 X--2 (covering two different wavelength ranges), with the characteristic emission lines included. The flux calibration was performed using a standard star (LTT7379) observed during the same period as the target. The spectrum displays a turnover in flux below $\sim 3800$ \AA, likely associated with a Balmer break that is likely explained by an A--type supergiant stellar companion to the ULX, thus providing the first (tentative) identification of the stellar type to this source. To highlight this, we overlay a sample of five best-fit stellar templates (solid lines) to the optical spectrum, with the optimised stellar parameters indicated in the legend. We redden the spectra according to a standard Milky Way diffuse law with $E(B-V) = 0.13 \pm 0.01$ and $R_v=3.1$, consistent with the method used in {\it{MUSE}} spectroscopy by \citet{zhou2022}. {\it{Bottom panel}}: Residuals to the best--fit with the colours corresponding to that of each stellar model in the upper panel.} 
\label{fig:masked_spec1D_ngc1313x2}
\end{figure*}
\begin{figure*}
\includegraphics[scale=0.5]{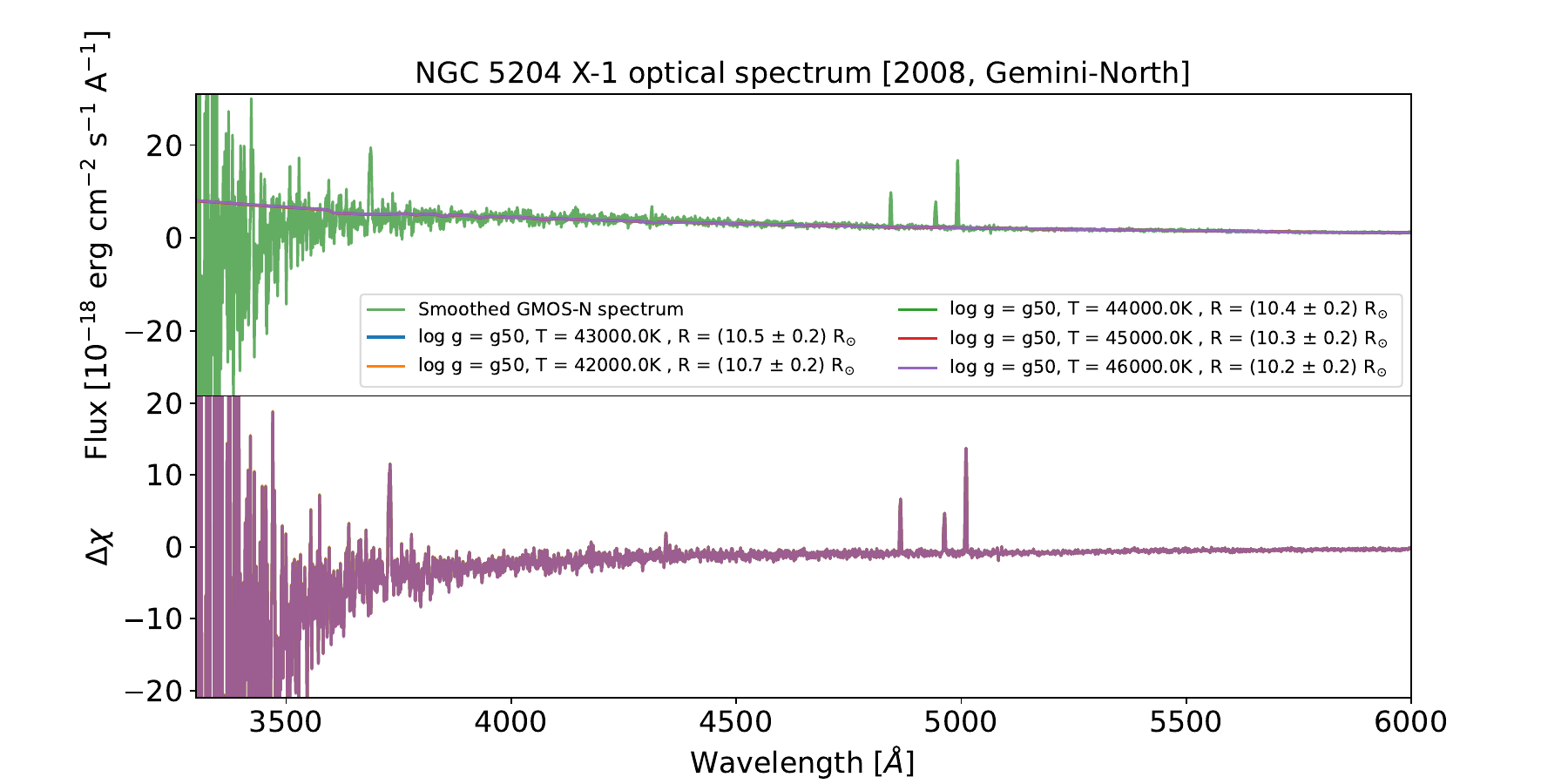}
\caption{A steeper {\it{Gemini}}--North (GMOS--N) optical spectrum of NGC 5204 X--1 reduced in the same manner as NGC 1313 X--2, with hot O/B type stellar models preferred over A--type models.}
\label{fig:masked_spec1D_ngc5204x1}
\end{figure*}
\begin{figure*}
\includegraphics[scale=0.5]{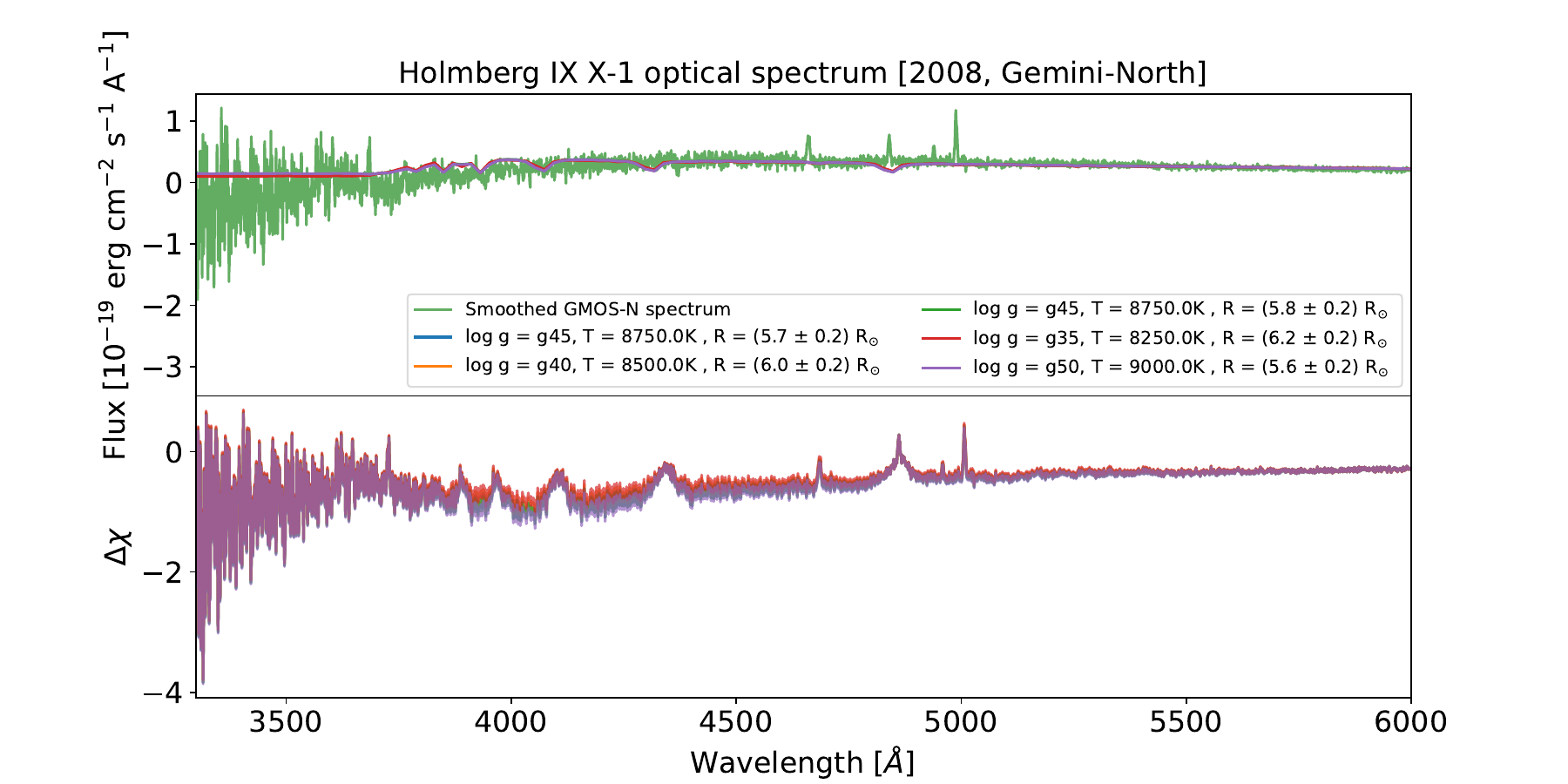}
\caption{Finally, the {\it{Gemini}}--North curved blackbody--like optical spectrum of Holmberg IX X--1, again distinct to NGC 1313 X--2. Here, the drop--off in wavelength below 4000 \AA is more easily distinguished.}
\label{fig:masked_spec1D_holmberg_ix_x1}
\end{figure*}
Spectral--timing studies have shown that the super--Eddington luminosities of many ULXs can be explained through a period of short lived high mass transfer rate from a Robe--lobe filling companion star onto a compact object or via wind--fed accretion from supergiant donors (\citealt{elmellah2019}). Indeed, the discovery of ultrafast winds in several ULXs through high resolution X--ray spectroscopy (\citealt{pinto2016}; \citealt{kosec2018}; \citealt{pinto2021}) and an anti-correlation between disc luminosity and temperature (\citealt{barra2022}; \citealt{robba2021}) provides evidence for accretion in these systems to be in excess of the Eddington limit and suggests that the inner disc emission is likely to be geometrically beamed. This would be expected in models of super--critical discs. Moreover, the detection of pulsations from as many as seven ULXs (\citealt{bachetti2014}; \citealt{israel2017a}; \citealt{fuerst16}; \citealt{israel2017b}; \citealt{carpano2018}; \citealt{sathyaprakash2019}; \citealt{rodriguezcastillo2019}; \citealt{vasilopoulos2020}; \citealt{kennea2017}) confirm the nature of the compact object to be neutron stars. Some of these sources show periodic Doppler shifts in their spin frequency, suggestive of orbital motion (\citealt{bachetti2014}). Many ULXs tend to be persistently bright, with their X--ray luminosity modulated over weekly timescales due to stochastic variations in the accretion rate, a warped precessing accretion flow (\citealt{brightman2019}), periodic eclipses (\citealt{urquhart2016}), or repeated interactions of a neutron star with the circumstellar disc of a Be--star (\citealt{Shigeyuki2022}). Some observations have reported occasional sharp drops in the X--ray luminosity, in accordance with the propeller effect, which has been known to be an indicator of the competition between the neutron star magnetic field and accretion rate (\citealt{tsygankov2016}; \citealt{earnshaw2018}; \citealt{earnshaw2024}).

Multi--wavelength observations of ULXs in the optical/UV and radio bands provide further information on the ULX environment, hinting an association in some cases with young star clusters aged between 20--30 Myrs (\citealt{grise09}; \citealt{grise11}; \citealt{grise12}), but have also revealed 300-$500$ pc bubble nebulae that are powered by photo--ionisation and shocks from disc winds (\citealt{pakull2006}; \citealt{pakull2008}). These nebulae could serve as calorimeters to constrain the kinematic and radiative energy budget of the ULX (\citealt{abolmasov2008}; \citealt{kaaret2010}). Recent works (\citealt{gurpide2024a}; \citealt{gurpide2024b}; \citealt{beuchert2024}) have used the photo--ionised emission line(s), including He II, alongside {\tt{Cloudy}} simulations to argue that these lines are insensitive to the hard ($> 2$ keV) X--ray emission from the inner disc regions and are thus poor indicators of the degree of geometrical beaming from the super--critical disc as a function of $L_{\text{X}}$, but can be used to constrain the EUV emission luminosity and the escape fraction $f_{\text{esc}}$. Indeed, in the case of NGC 1313 X--2, \citet{gurpide2024a} placed an upper limit on the He II 4686 emission line luminosity from the bubble nebula to be $< 10^{36}$ erg s$^{-1}$ with high resolution MUSE data (see Figure 3 of their paper), quoting an EUV luminosity $\lesssim 10^{39}$ erg s$^{-1}$, after considering the effects of ISM density, abundances, filling factor and covering factor of the nebula. By comparison, this was found to be smaller than the EUV luminosity of NGC 6946 X--1 by an order of magnitude, likely implying a weaker wind and a smaller mass transfer rate for NGC 1313 X--2. 

In the case of Holmberg II X-1 (\citealt{cseh2014}), NGC 7793 S26 (\citealt{pakull2010}) and NGC 2276-3c (\citealt{mezcua2015}), VLBI radio data resolve the nebulae into a core and adjacent lobes, characteristic of being powered by collimated, synchrotron dominated jets that are transient, possibly similar to that observed in near--Eddington accretion states of Galactic X--ray binaries, in a similar vein to some sub--populations within the super--Eddington accretion regime (\citealt{Koljonen2021}).     
Data from optical and UV bands have also been used to characterise the emission components expected to be present in regions more local to the ULX, with the aim of determining the spectral type, mass and metallicity of the donor star in the binary, in addition to the orbital parameters, by computing radial velocities, providing an insight into the formation and evolution of ULXs with population synthesis models. For example, deep spectroscopic observations of the pulsating ULX NGC 7793 P13 with {\it{VLT}} and {\it{Swift}} UVOT (\citealt{motch2014}) revealed a B9Ia supergiant star owing to the presence of Mg [II, $\lambda$ 4481], Si [II, $\lambda$ 4128--4130] and Fe [II] absorption lines in the optical spectrum. Such campaigns are hence useful to constrain the compact object mass. 

A more detailed picture of the system, including the inclination, eccentricity, orbital and super-orbital periods, were obtained by modelling the phase delays between the u--band and V--band modulations with an eclipsing light--curve model that assumes a precessing disc casting shadows on an X--ray heated stellar atmosphere. Other examples in which the spectral type of the donor have been reliably confirmed are M101 ULX-1 (\citealt{Liu2013}) and NGC 300 ULX-1 (\citealt{heida2019}). However, obtaining sufficiently high quality data to perform such analyses for a large sample of ULXs is challenging due to their large distances, sometimes resulting in multiple optical counterparts to a given X--ray source. The advent of high--resolution imaging with {\it{HST}}, and more recently with {\it{JWST}}, has enabled a precise localisation of unique optical and infrared counterparts to several ULXs (\citealt{roberts2008}; \citealt{gladstone2013}; \citealt{allak2024}).      

In other ULXs, the observed hourly--weekly timescale variability in the optical (V and B) bands implies evidence for disc--dominated or wind--dominated emission (\citealt{Tao2011}). Disentangling the stellar contribution from these additional emission components remains a challenge, but has been attempted by modelling the spectral energy distribution (SED) of ULXs with irradiated disc models. Such models can provide more realistic constraints on the accretion luminosity, the disc size and geometry (see e.g. ESO 243-49 HLX-1; \citealt{soria2017}; \citealt{webb2017}). Alternatively, infrared observations of ULX counterparts are largely unaffected by disc reprocessing, and have been useful in uncovering a number of red supergiant donor star candidates, initially with imaging studies from {\it{WHT/LIRIS}}, {\it{VLT}} and {\it{MMT}} (\citealt{heida2014}), followed by high resolution spectroscopic campaigns with Keck/MOSFIRE (\citealt{heida2016}; \citealt{heida2019}). However, results from \cite{lopez2017,lopez2020} and \cite{lau2019} suggest additional components to the infrared emission in ULXs, with a mid--IR excess in some sources pointing to emission from a circumbinary disc owing to material lost through the L2 Lagrangian point (see \citealt{dudik2016}), and evidence of variable infrared emission potentially suggesting the de--brightening phase of a synchrotron dominated jet (\citealt{lau2019}; \citealt{sathyaprakash2022}). However, degeneracies between the different models remain due to the non--simultaneity of the X--ray and infrared observations. In this work, we report on twelve nights of {\it{Gemini-South}} (GMOS-S) observations of NGC 1313 X--2 using archival data from 2009 to constrain the stellar type of the donor star, discuss the significance of a marginal two day delay between the X--ray and optical emission, keeping in mind the resulting implications.  

\subsection{NGC 1313 X-2}

The ULX NGC 1313 X--2, member of a barred spiral galaxy and located at a distance of $D=4.13$ Mpc (calibrated with a population of classical Type I Cepheids) was the target of several X--ray and optical observations. The formation history of the ULX, which is closely tied to the formation of the central compact object, remains debated. {\it{HST}} images show the optical counterpart to reside well outside the D25 isophotal ellipse towards the southern region of the galaxy (i.e. $\sim$ 10 kpc from the nucleus), near areas that lack on--going star--formation activity. However, a young O or B type stellar association is also found adjacent to the ULX, alongside an otherwise (predominantly) old and intermediate--age population (\citealt{grise09}). However, the O/B assocation does not reside in the spiral arms of the galaxy and was hence posited to originate due to a localised starburst event. Such a starburst episode was thought to have been triggered by a tidal interaction with a satellite galaxy, or more likely, with a fast--moving HI cloud transiting across the disc plane.  
\newline
\newline
X--ray observations provide further insight into its emission properties (\citealt{zampieri_2003}), and historically, the disc emission was thought to arise from a sub--Eddington accretion disc. However, one was forced towards an alternative hypothesis by the phenomenological high energy cut-off in the {\it{NuSTAR}} data. The ULX has a peak luminosity $\sim 6 \times 10^{39}$ erg s$^{-1}$, and long--term {\it{Swift}}--XRT monitoring indicates super--orbital periodicity, with a period of $\sim 25$ days, in a similar vein to other well studied ULXs. More importantly, pulsations were clearly discovered in this source (\citealt{sathyaprakash2019}) providing more weight to the nature of source as a neutron star rather than an IMBH. 

\section{Data reduction}

\begin{figure}
\includegraphics[scale=0.5]{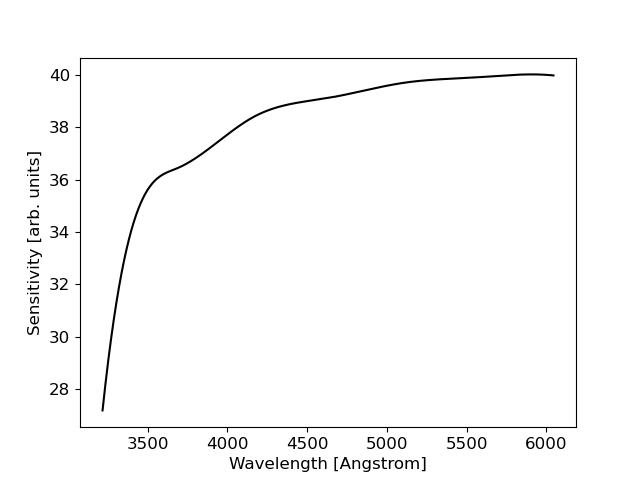}
\caption{The sensitivity curve used for converting instrumental {\it{Gemini}}--South count rates of NGC 1313 X--2 to flux in standard units (erg cm$^{-2}$ s$^{-1}$ \AA$^{-1}$.}
\label{fig:sens_curve}
\end{figure}

\begin{table*}
\centering
\caption{Summary of {\it{Gemini-South}} observations of NGC 1313 X--2.}
\label{tab:gemini_observations}
\begin{tabular}{ccccc}
\hline
\hline
Observation night & Observation identifier & Central wavelength [\AA] & Airmass & Exposure time [s] \\
\hline
\hline
2005-10-08 & GS-2005B-Q-30-1-002 & 455 & 1.25 & 1200.47 \\
2005-10-08 & GS-2005B-Q-30-1-003 & 455 & 1.24 & 1200.47 \\
2005-10-08 & GS-2005B-Q-30-1-004 & 455 & 1.24 & 1200.47 \\
2005-10-08 & GS-2005B-Q-30-1-005 & 460 & 1.25 & 1200.47 \\
2005-10-08 & GS-2005B-Q-30-1-006 & 460 & 1.26 & 1200.47 \\
2005-10-08 & GS-2005B-Q-30-1-007 & 460 & 1.27 & 1200.47 \\
2005-10-08 & GS-2005B-Q-30-1-009 & 465 & 1.29 & 1200.47 \\
2005-10-08 & GS-2005B-Q-30-1-010 & 465 & 1.31 & 1200.47 \\
2005-10-08 & GS-2005B-Q-30-1-011 & 465 & 1.34 & 1200.47 \\
2005-11-30 & GS-2005B-Q-30-1-013 & 460 & 1.24 & 1200.47 \\
2005-11-30 & GS-2005B-Q-30-1-014 & 460 & 1.25 & 1200.47 \\
\hline
\hline
2009-10-23 & GS-2009B-Q-20-1-001 & 456 & 1.246 & 918.48 \\
2009-10-23 & GS-2009B-Q-20-1-002 & 456 & 1.241 & 918.48 \\
2009-10-23 & GS-2009B-Q-20-1-003 & 456 & 1.239 & 918.48 \\
2009-10-23 & GS-2009B-Q-20-1-004 & 456 & 1.239 & 3.50 \\
2009-10-23 & GS-2009B-Q-20-1-005 & 461 & 1.239 & 3.50 \\
2009-10-23 & GS-2009B-Q-20-1-006 & 461 & 1.240 & 918.48 \\
2009-10-23 & GS-2009B-Q-20-1-007 & 461 & 1.244 & 918.48 \\
2009-10-23 & GS-2009B-Q-20-1-008 & 461 & 1.250 & 918.48 \\
2009-10-23 & GS-2009B-Q-20-1-009 & 461 & 1.259 & 918.48 \\
2009-10-23 & GS-2009B-Q-20-1-010 & 466 & 1.271 & 918.48 \\
2009-10-23 & GS-2009B-Q-20-1-011 & 466 & 1.286 & 918.48 \\
2009-10-23 & GS-2009B-Q-20-1-012 & 466 & 1.303 & 918.48 \\
2009-10-23 & GS-2009B-Q-20-1-013 & 466 & 1.314 & 3.50 \\
2009-12-11 & GS-2009B-Q-20-2-001 & 456 & 1.256 & 918.48 \\
2009-12-11 & GS-2009B-Q-20-2-002 & 456 & 1.267 & 918.48 \\
2009-12-11 & GS-2009B-Q-20-2-003 & 456 & 1.280 & 918.48 \\
2009-12-11 & GS-2009B-Q-20-2-004 & 456 & 1.289 & 3.50 \\
2009-12-11 & GS-2009B-Q-20-2-005 & 461 & 1.290 & 3.50 \\
2009-12-11 & GS-2009B-Q-20-2-006 & 461 & 1.300 & 918.48 \\
2009-12-11 & GS-2009B-Q-20-2-007 & 461 & 1.320 & 918.48 \\
2009-12-11 & GS-2009B-Q-20-2-008 & 461 & 1.343 & 918.48 \\
2009-12-11 & GS-2009B-Q-20-2-009 & 461 & 1.370 & 918.48 \\
2009-12-11 & GS-2009B-Q-20-2-010 & 466 & 1.401 & 918.48 \\
2009-12-11 & GS-2009B-Q-20-2-011 & 466 & 1.435 & 918.48 \\
2009-12-11 & GS-2009B-Q-20-2-012 & 466 & 1.474 & 918.48 \\
2009-12-11 & GS-2009B-Q-20-2-013 & 466 & 1.496 & 3.50 \\
2009-12-12 & GS-2009B-Q-20-3-001 & 456 & 1.243 & 918.48 \\
2009-12-12 & GS-2009B-Q-20-3-002 & 456 & 1.240 & 918.48 \\
2009-12-12 & GS-2009B-Q-20-3-003 & 456 & 1.239 & 918.48 \\
2009-12-12 & GS-2009B-Q-20-3-004 & 456 & 1.240 & 3.50 \\
2009-12-12 & GS-2009B-Q-20-3-005 & 461 & 1.240 & 3.50 \\
2009-12-12 & GS-2009B-Q-20-3-006 & 461 & 1.242 & 918.48 \\
2009-12-12 & GS-2009B-Q-20-3-007 & 461 & 1.247 & 918.48 \\
2009-12-12 & GS-2009B-Q-20-3-008 & 461 & 1.255 & 918.48 \\
2009-12-12 & GS-2009B-Q-20-3-009 & 461 & 1.265 & 918.48 \\
2009-12-12 & GS-2009B-Q-20-3-010 & 466 & 1.279 & 918.48 \\
2009-12-12 & GS-2009B-Q-20-3-011 & 466 & 1.295 & 918.48 \\
2009-12-12 & GS-2009B-Q-20-3-012 & 466 & 1.314 & 918.48 \\
2009-12-12 & GS-2009B-Q-20-3-013 & 466 & 1.326 & 3.50 \\
\hline
\end{tabular}
\end{table*}

\begin{table*}
\centering
\caption{Summary of {\it{Gemini-South}} observations for NGC 1313 X-2 (continued)}
\begin{tabular}{ccccc}
\hline
\hline
Observation night & Observation identifier & Central wavelength [\AA] & Airmass & Exposure time [s] \\
\hline
\hline
2009-12-13 & GS-2009B-Q-20-4-001 & 456 & 1.276 & 918.48 \\
2009-12-13 & GS-2009B-Q-20-4-002 & 456 & 1.291 & 918.48 \\
2009-12-13 & GS-2009B-Q-20-4-003 & 456 & 1.310 & 918.48 \\
2009-12-13 & GS-2009B-Q-20-4-004 & 456 & 1.321 & 3.50 \\
2009-12-13 & GS-2009B-Q-20-4-005 & 461 & 1.323 & 3.50 \\
2009-12-13 & GS-2009B-Q-20-4-006 & 461 & 1.335 & 918.48 \\
2009-12-13 & GS-2009B-Q-20-4-007 & 461 & 1.361 & 918.48 \\
2009-12-13 & GS-2009B-Q-20-4-008 & 461 & 1.390 & 918.48 \\
2009-12-13 & GS-2009B-Q-20-4-009 & 461 & 1.423 & 918.48 \\
2009-12-13 & GS-2009B-Q-20-4-010 & 466 & 1.461 & 918.48 \\
2009-12-13 & GS-2009B-Q-20-4-011 & 466 & 1.502 & 918.48 \\
2009-12-13 & GS-2009B-Q-20-4-012 & 466 & 1.548 & 918.48 \\
2009-12-13 & GS-2009B-Q-20-4-013 & 466 & 1.575 & 3.50 \\
2009-12-14 & GS-2009B-Q-20-5-001 & 456 & 1.300 & 918.48 \\
2009-12-14 & GS-2009B-Q-20-5-002 & 456 & 1.320 & 918.48 \\
2009-12-14 & GS-2009B-Q-20-5-003 & 456 & 1.344 & 918.48 \\
2009-12-14 & GS-2009B-Q-20-5-004 & 456 & 1.358 & 3.50 \\
2009-12-14 & GS-2009B-Q-20-5-005 & 461 & 1.360 & 3.50 \\
2009-12-14 & GS-2009B-Q-20-5-006 & 461 & 1.375 & 918.48 \\
2009-12-14 & GS-2009B-Q-20-5-007 & 461 & 1.406 & 918.48 \\
2009-12-14 & GS-2009B-Q-20-5-008 & 461 & 1.441 & 918.48 \\
2009-12-14 & GS-2009B-Q-20-5-009 & 461 & 1.480 & 918.48 \\
2009-12-14 & GS-2009B-Q-20-5-010 & 466 & 1.525 & 918.48 \\
2009-12-14 & GS-2009B-Q-20-5-011 & 466 & 1.574 & 918.48 \\
2009-12-14 & GS-2009B-Q-20-5-012 & 466 & 1.628 & 918.48 \\
2009-12-14 & GS-2009B-Q-20-5-013 & 466 & 1.660 & 3.50 \\
2009-12-15 & GS-2009B-Q-20-6-001 & 456 & 1.254 & 918.48 \\
2009-12-15 & GS-2009B-Q-20-6-002 & 456 & 1.264 & 918.48 \\
2009-12-15 & GS-2009B-Q-20-6-003 & 456 & 1.277 & 918.48 \\
2009-12-15 & GS-2009B-Q-20-6-004 & 456 & 1.285 & 3.50 \\
2009-12-15 & GS-2009B-Q-20-6-005 & 461 & 1.286 & 3.50 \\
2009-12-15 & GS-2009B-Q-20-6-006 & 461 & 1.295 & 918.48 \\
2009-12-15 & GS-2009B-Q-20-6-007 & 461 & 1.314 & 918.48 \\
2009-12-15 & GS-2009B-Q-20-6-008 & 461 & 1.337 & 918.48 \\
2009-12-15 & GS-2009B-Q-20-6-009 & 461 & 1.363 & 918.48 \\
2009-12-15 & GS-2009B-Q-20-6-010 & 466 & 1.393 & 918.48 \\
2009-12-15 & GS-2009B-Q-20-6-011 & 466 & 1.426 & 918.48 \\
2009-12-15 & GS-2009B-Q-20-6-012 & 466 & 1.463 & 918.48 \\
2009-12-15 & GS-2009B-Q-20-6-013 & 466 & 1.485 & 3.50 \\
2009-12-16 & GS-2009B-Q-20-7-001 & 456 & 1.314 & 918.48 \\
2009-12-16 & GS-2009B-Q-20-7-002 & 456 & 1.337 & 918.48 \\
2009-12-16 & GS-2009B-Q-20-7-003 & 456 & 1.363 & 918.48 \\
2009-12-16 & GS-2009B-Q-20-7-004 & 456 & 1.378 & 3.50 \\
2009-12-16 & GS-2009B-Q-20-7-005 & 461 & 1.380 & 3.50 \\
2009-12-16 & GS-2009B-Q-20-7-006 & 461 & 1.397 & 918.48 \\
2009-12-16 & GS-2009B-Q-20-7-007 & 461 & 1.431 & 918.48 \\
\hline
\end{tabular}
\end{table*}

\begin{table*}
\centering
\caption{Summary of {\it{Gemini-South}} observations for NGC 1313 X-2 (continued)}
\begin{tabular}{ccccc}
\hline
\hline
Observation night & Observation identifier & Central wavelength [\AA] & Airmass & Exposure time [s] \\
\hline
\hline
2009-12-20 & GS-2009B-Q-20-8-001 & 456 & 1.258 & 918.48 \\
2009-12-20 & GS-2009B-Q-20-8-002 & 456 & 1.250 & 918.48 \\
2009-12-20 & GS-2009B-Q-20-8-003 & 456 & 1.246 & 918.48 \\
2009-12-20 & GS-2009B-Q-20-8-004 & 456 & 1.241 & 3.50 \\
2009-12-20 & GS-2009B-Q-20-8-005 & 461 & 1.240 & 3.50 \\
2009-12-20 & GS-2009B-Q-20-8-006 & 461 & 1.241 & 918.48 \\
2009-12-20 & GS-2009B-Q-20-8-007 & 461 & 1.244 & 918.48 \\
2009-12-20 & GS-2009B-Q-20-8-008 & 461 & 1.251 & 918.48 \\
2009-12-20 & GS-2009B-Q-20-8-009 & 461 & 1.260 & 918.48 \\
2009-12-20 & GS-2009B-Q-20-8-010 & 466 & 1.272 & 918.48 \\
2009-12-20 & GS-2009B-Q-20-8-011 & 466 & 1.288 & 918.48 \\
2009-12-20 & GS-2009B-Q-20-8-012 & 466 & 1.308 & 918.48 \\
2009-12-20 & GS-2009B-Q-20-8-013 & 466 & 1.320 & 3.50 \\
2009-12-21 & GS-2009B-Q-20-9-001 & 456 & 1.252 & 918.48 \\
2009-12-21 & GS-2009B-Q-20-9-002 & 456 & 1.262 & 918.48 \\
2009-12-21 & GS-2009B-Q-20-9-003 & 456 & 1.276 & 918.48 \\
2009-12-21 & GS-2009B-Q-20-9-004 & 456 & 1.284 & 3.50 \\
2009-12-21 & GS-2009B-Q-20-9-005 & 461 & 1.286 & 3.50 \\
2009-12-21 & GS-2009B-Q-20-9-006 & 461 & 1.298 & 918.48 \\
2009-12-21 & GS-2009B-Q-20-9-007 & 461 & 1.320 & 918.48 \\
2009-12-21 & GS-2009B-Q-20-9-008 & 461 & 1.344 & 918.48 \\
2009-12-21 & GS-2009B-Q-20-9-009 & 461 & 1.374 & 918.48 \\
2009-12-21 & GS-2009B-Q-20-9-010 & 466 & 1.409 & 918.48 \\
2009-12-21 & GS-2009B-Q-20-9-011 & 466 & 1.448 & 918.48 \\
2009-12-21 & GS-2009B-Q-20-9-012 & 466 & 1.492 & 918.48 \\
2009-12-21 & GS-2009B-Q-20-9-013 & 466 & 1.516 & 3.50 \\
2009-12-22 & GS-2009B-Q-20-10-001 & 456 & 1.250 & 918.48 \\
2009-12-22 & GS-2009B-Q-20-10-002 & 456 & 1.259 & 918.48 \\
2009-12-22 & GS-2009B-Q-20-10-003 & 456 & 1.272 & 918.48 \\
2009-12-22 & GS-2009B-Q-20-10-004 & 456 & 1.282 & 3.50 \\
2009-12-22 & GS-2009B-Q-20-10-005 & 461 & 1.284 & 3.50 \\
2009-12-22 & GS-2009B-Q-20-10-006 & 461 & 1.297 & 918.48 \\
2009-12-22 & GS-2009B-Q-20-10-007 & 461 & 1.319 & 918.48 \\
2009-12-22 & GS-2009B-Q-20-10-008 & 461 & 1.344 & 918.48 \\
2009-12-22 & GS-2009B-Q-20-10-009 & 461 & 1.374 & 918.48 \\
2009-12-22 & GS-2009B-Q-20-10-010 & 466 & 1.409 & 918.48 \\
2009-12-22 & GS-2009B-Q-20-10-011 & 466 & 1.448 & 918.48 \\
2009-12-22 & GS-2009B-Q-20-10-012 & 466 & 1.492 & 918.48 \\
2009-12-22 & GS-2009B-Q-20-10-013 & 466 & 1.517 & 3.50 \\
2009-12-24 & GS-2009B-Q-20-42-001 & 456 & 1.243 & 918.48 \\
2009-12-24 & GS-2009B-Q-20-42-002 & 456 & 1.240 & 918.48 \\
2009-12-24 & GS-2009B-Q-20-42-003 & 456 & 1.239 & 918.48 \\
2009-12-24 & GS-2009B-Q-20-42-004 & 456 & 1.240 & 3.50 \\
2009-12-24 & GS-2009B-Q-20-42-005 & 461 & 1.240 & 3.50 \\
2009-12-24 & GS-2009B-Q-20-42-006 & 461 & 1.247 & 918.48 \\
2009-12-24 & GS-2009B-Q-20-42-007 & 461 & 1.254 & 918.48 \\
2009-12-24 & GS-2009B-Q-20-42-008 & 461 & 1.265 & 918.48 \\
2009-12-24 & GS-2009B-Q-20-42-009 & 461 & 1.278 & 918.48 \\
2009-12-24 & GS-2009B-Q-20-42-010 & 466 & 1.294 & 918.48 \\
2009-12-24 & GS-2009B-Q-20-42-011 & 466 & 1.314 & 918.48 \\
2009-12-24 & GS-2009B-Q-20-42-012 & 466 & 1.336 & 918.48 \\
2009-12-24 & GS-2009B-Q-20-42-013 & 466 & 1.349 & 3.50 \\
\hline
\end{tabular}
\end{table*}
The main characteristics of the data used in this study are summarised in Table~\ref{tab:gemini_observations}. The target was observed with the {\it{Gemini-South}} observatory located in the Chilean Andes over ten nights between October--December 2009, yielding a total of ten science frames per night, each with an exposure of 918 seconds. Some observations were also scheduled during August 2005. Long--slit spectra were obtained using a 1.0 arc--second slit with a B600 grating: for more details please see the observation log Tables~\ref{tab:gemini_observations}. In order to mitigate the effect of gaps in the GMOS-S detectors, the science frames were observed in a three-position dither pattern, corresponding to central wavelengths $\lambda_c$={456,461,466} nm, providing spectral coverage between 3500--6000 Angstrom. On average, three science frames corresponding to a given dither pattern were available per night, in addition to the associated flat--field frames, bias frames and arc lamps, which were used for wavelength calibration. 

We adopted routines within the {\tt{pyraf}} package (\citealt{pyraf2012}) to perform the data reduction, which mainly includes wavelength and flux calibrations. To convert the spectrum from counts s$^{-1}$ to a flux density (i.e. flux calibration), we used GMOS--S observations of the standard star LTT7379 observed over the same wavelength range as the source, with its prior spectrum taken from the {\tt{onedspec}} database in {\tt{pyraf}}. In order to subtract the mean bias level, which is applied to mitigate electronic read--out noise from the science frames and calibration frames, we co--added bias frames for each observing night using the {\tt{gbias}} routine in {\tt{pyraf}}. We then flat--fielded the science frames and arc frames to correct for a non-uniform sensitivity of the GMOS pixels for a given dither configuration with {\tt{gsflat}}. We model the dependence of the GMOS pixel sensitivity as a function of wavelength with a cubic spline. We then apply wavelength calibration to the science frames using {\tt{gstransform}} after identifying emission lines from the CuAr arc lamp frames with the {\tt{gswavelength}} routine. We ensured that the derived wavelength solution was accurate to within 0.5 Angstrom. Finally, we subtracted the background from each science frame by sampling the median background level in two regions above and below the main target, while rejecting pixels deviating from the median by more than 3$\sigma$. Finally, the sky subtracted science frames for the three dither configurations from all twelve nights were co--added with {\tt{gemcombine}} using median stacking and $3\sigma$ clipping to exclude pixels affected by cosmic ray hits. In order to reduce the standard star observations, we adopted the same procedure as described above. We then extracted a 1D spectrum of the target and the standard star from the stacked 2D science frames using a suitably chosen aperture with {\tt{apall}}. A sensitivity curve was derived using {\tt{sensfunc}} and flux calibration was applied using {\tt{calibrate}}. The entire analysis with {\tt{pyraf}} was run on a virtual machine \footnote{https://gemini-iraf-vm-tutorial.readthedocs.io/en/stable/index.html}.

We present the stacked 2D science frame of the source alongside the 1D spectrum in Figures~\ref{fig:stacked_spec2D} and~\ref{fig:masked_spec1D_ngc1313x2}, and the location of the optical counterpart in Figure~\ref{fig:galaxy_image_ngc1313}. The stacked spectrum was smoothed with a Savitzky-Golay filter (with a window length of 40 Angstrom and polynomial order 4) to remove high frequency noise.  

\section{Results}

\begin{table*}
\caption{Best--fit parameters corresponding to the observed emission line profiles modelled using a combination of Lorentzian and Gaussian functions.}
\centering	
\begin{tabular}{rrrrrr}
\hline
\hline
Emission line & Epoch [MJD] & $\lambda_{c}$ [Angstrom] & FWHM [Angstrom] & Velocity [km s$^{-1}$] & Velocity dispersion [km s$^{-1}$] \\
\hline
\hline
[OIII] 5007 & 55127.0 & 5011.7 $\pm$ 0.1 & 2.6 $\pm$ 0.03 & -180 $\pm$ 2 & 307 $\pm$ 3 \\
& 55176.0 & 5011.9 $\pm$ 0.1 & 3.00 $\pm$ 0.03 & -170 $\pm$ 2 & 358 $\pm$ 4 \\
& 55177.0 & 5011.4 $\pm$ 0.1 & 2.95 $\pm$ 0.03 & -195 $\pm$ 2 & 353 $\pm$ 4 \\
& 55178.0 & 5011.7 $\pm$ 0.1 & 2.55 $\pm$ 0.03 & -180 $\pm$ 2 & 305 $\pm$ 3 \\
& 55179.0 & 5011.8 $\pm$ 0.1 & 2.90 $\pm$ 0.03 & -170 $\pm$ 2 & 347 $\pm$ 3 \\
& 55180.0 & 5012.1 $\pm$ 0.1 & 2.43 $\pm$ 0.03 & -155 $\pm$ 1 & 290 $\pm$ 2 \\
& 55185.0 & 5011.9 $\pm$ 0.1 & 2.01 $\pm$ 0.03 & -165 $\pm$ 2 & 241 $\pm$ 4 \\
& 55186.0 & 5011.8 $\pm$ 0.1 & 2.65 $\pm$ 0.03 & -170 $\pm$ 2 & 317 $\pm$ 4 \\
& 55187.0 & 5012.0 $\pm$ 0.1 & 2.99 $\pm$ 0.04 & -160 $\pm$ 2 & 357 $\pm$ 3 \\
& 55189.0 & 5011.7 $\pm$ 0.1 & 2.68 $\pm$ 0.04 & -180 $\pm$ 2 & 320 $\pm$ 4 \\
\hline
[OIII] 4959 & 55127.0 & 4963.6 $\pm$ 0.1 & 4.7 $\pm$ 0.1 & -190 $\pm$ 5 & 470 $\pm$ 30 \\
& 55176.0 & 4963.7 $\pm$ 0.1 & 2.0 $\pm$ 0.1 & -180 $\pm$ 7 & 330 $\pm$ 40 \\
& 55177.0 & 4963.4 $\pm$ 0.1 & 2.1 $\pm$ 0.1 & -200 $\pm$ 6 & 380 $\pm$ 30 \\
& 55178.0 & 4963.9 $\pm$ 0.1 & 4.0 $\pm$ 0.1 & -170 $\pm$ 5 & 440 $\pm$ 50 \\
& 55179.0 & 4964.0 $\pm$ 0.1 & 3.1 $\pm$ 0.1 & -160 $\pm$ 5 & 550 $\pm$ 30 \\
& 55180.0 & 4964.1 $\pm$ 0.1 & 1.9 $\pm$ 0.1 & -155 $\pm$ 3 & 400 $\pm$ 40 \\
& 55185.0 & 4963.7 $\pm$ 0.1 & 3.1 $\pm$ 0.1 & -180 $\pm$ 6 & $<1000$ \\
& 55186.0 & 4963.7 $\pm$ 0.1 & 2.6 $\pm$ 0.1 & -180 $\pm$ 7 & 500 $\pm$ 10 \\
& 55187.0 & 4963.6 $\pm$ 0.1 & 2.8 $\pm$ 0.1 & -190 $\pm$ 6 & 490 $\pm$ 10 \\
& 55189.0 & 4963.7 $\pm$ 0.1 & 2.9 $\pm$ 0.1 & -180 $\pm$ 6 & 400 $\pm$ 20 \\
\hline
[He II] 4686 & 55127.0 & 4689.7 $\pm$ 0.6 & 6.4 $\pm$ 0.1 & -210 $\pm$ 10 & 320 $\pm$ 20 \\
& 55176.0 & 4692.7 $\pm$ 0.5 & 3.3 $\pm$ 0.1 & -23 $\pm$ 2 & 430 $\pm$ 10 \\
& 55177.0 & 4689.8 $\pm$ 0.4 & 3.3 $\pm$ 0.1 & -210 $\pm$ 20 & 420 $\pm$ 10 \\
& 55178.0 & 4690.9 $\pm$ 0.4 & 2.3 $\pm$ 0.1 & -140 $\pm$ 20 & 310 $\pm$ 30 \\
& 55179.0 & 4692.2 $\pm$ 0.9 & 3.3 $\pm$ 0.1 & -55 $\pm$ 40 & 400 $\pm$ 50 \\
& 55180.0 & 4690.7 $\pm$ 1 & 1.3 $\pm$ 0.1 & -150 $\pm$ 50 & 175 $\pm$ 30 \\
& 55185.0 & 4698.1 $\pm$ 0.3 & 120 $\pm$ 0.1 & -480  $\pm$ 30 & 150 $\pm$ 40 \\
& 55186.0 & 4688.7 $\pm$ 0.8 & 1.8 $\pm$ 0.1 & 390  $\pm$ 30 & $<10$ \\
& 55187.0 & 4693.0 $\pm$ 0.1 & -0.9 $\pm$ 0.1 & -620 $\pm$ 40 & 50 $\pm$ 5 \\
& 55189.0 & 4688.4 $\pm$ 0.4 & 3.0 $\pm$ 0.1 & -290 $\pm$ 20 & 300 $\pm$ 10 \\
\hline
\hline
\end{tabular}
\label{Tab:emission_line_fits}
\end{table*}

\begin{table*}
\caption{Optimised parameters corresponding to the stellar models fit to the stacked {\it{Gemini--South}} spectrum of NGC 1313 X--2. We quote the best--fit surface gravity, radius, surface temperature, metal abundance and the reduced $\chi^{2}$.}
\centering	
\begin{tabular}{rrrrrr}
\hline
\hline
Stellar model & $\log g$ & Radius [$R_{\odot}$] & $T_{\text{eff}}$ [K] & [M/H] & $\chi^{2}$/dof \\
\hline
1 & +1.0 & 47.1 $\pm 0.2$ & 8250 & +0.5 & 7788/5554 \\
2 & +1.0 & 55.9 $\pm 0.2$ & 8250 & -0.5 & 7825/5554 \\
3 & +1.0 & 47.4 $\pm 0.2$ & 8250 & +0.2 & 7846/5554 \\
4 & +1.0 & 55.6 $\pm 0.2$ & 8250 & +0.0 & 7907/5554 \\
5 & +1.0 & 47.6 $\pm 0.2$ & 8000 & +0.0 & 7927/5554 \\
6 & +1.0 & 56.1 $\pm 0.2$ & 7500 & -1.0 & 7957/5554 \\
7 & +0.5 & 55.5 $\pm 0.2$ & 7500 & +0.2 & 7964/5554 \\
8 & +1.0 & 48.9 $\pm 0.2$ & 8000 & +0.5 & 7968/5554 \\
9 & +0.5 & 48.0 $\pm 0.2$ & 8250 & -0.5 & 7974/5554 \\
10 & +1.0 & 56.3 $\pm 0.2$ & 7500 & -1.5 & 7986/5554 \\
\hline
\hline
\end{tabular}
\label{Tab:stellar_fits}
\end{table*}

\begin{figure*}
\includegraphics[scale=0.5]{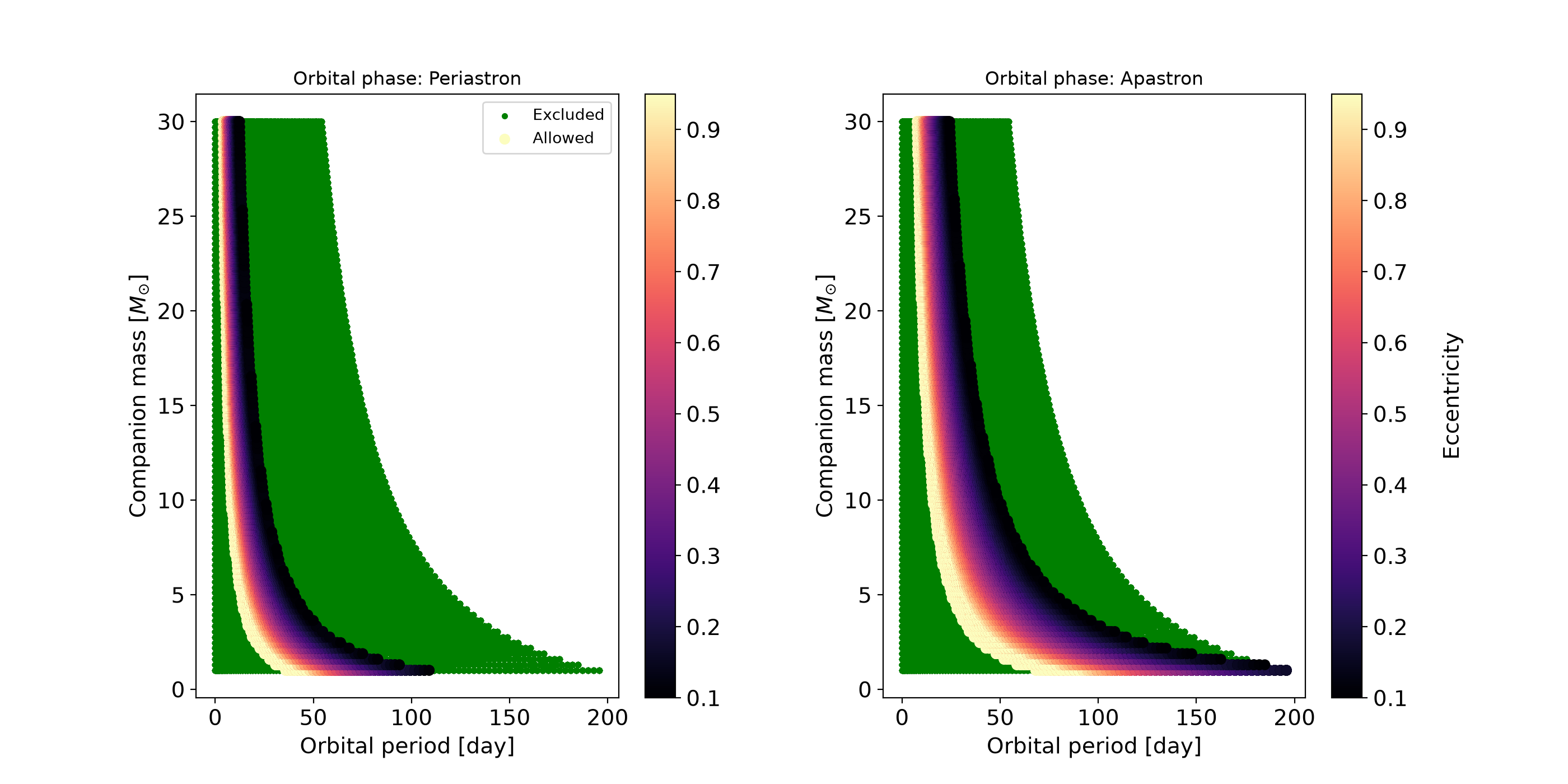}
\caption{Constraints on the orbital parameters including the orbital separation, companion mass and eccentricity of NGC 1313 X--2, assuming that the mass transfer in this system occurs through Roche--Lobe overflow, such that the stellar radius is limited to $\lesssim$ Roche-Lobe radius (see text). The Roche--Lobe radius was computed for two extreme orbital phases: i.e. at the periastron and apastron to indicate the full range of allowed orbital solutions. The colour--bar maps the eccentricity, while the orbital period and companion mass are included in the horizontal and vertical axes. The parameters that are excluded by the above condition are shown in green.}
\label{fig:orb_constraints}
\end{figure*}

\begin{figure*}
\includegraphics[scale=0.5]{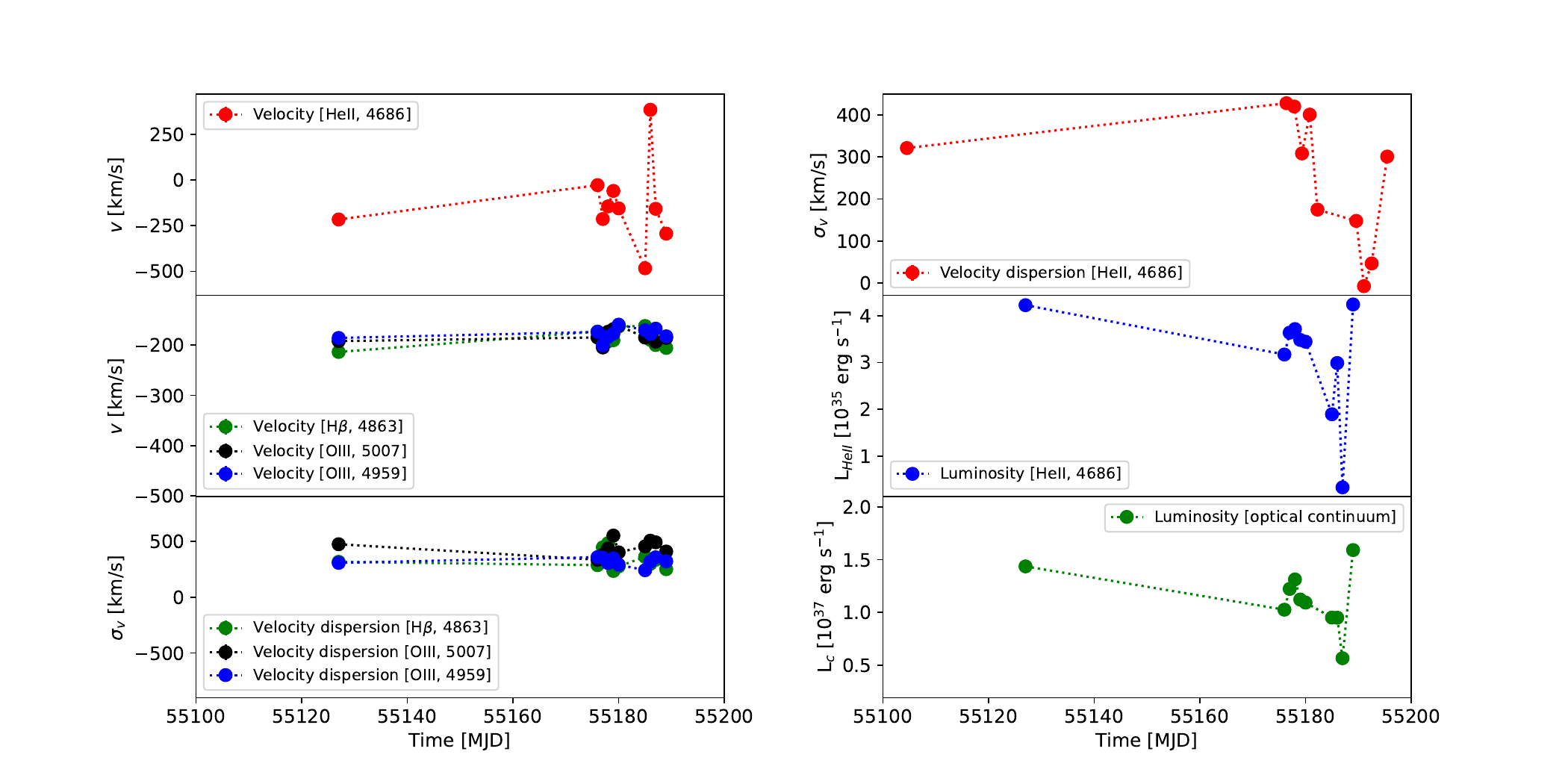}
\caption{Here, we show the time evolution of the radial velocities $v$ and velocity dispersions $\sigma_{v}$ of (i) the [He II, $\lambda 4686$] emission line and (ii) transitions that occur within the shock--ionised nebula including [H$\beta$, 4863], [OIII, 4959] and [OIII, 5007]. Clearly, radial velocities of the bubble indicates that it expands at a more steady rate compared to the dynamical motions of the ULX. We also plot the [He II, $\lambda 4686$] luminosity and the optical continuum luminosity as a function of epoch. We infer the radial velocity of the nebula by fitting the associated emission lines with a Voigt profile and subtract the systemic velocity of the host galaxy (see Figures~\ref{fig:nebular_lines1},~\ref{fig:nebular_lines2} and~\ref{fig:nebular_lines3}). We note that a negative velocity value indicates blue--shifted emission.}
\label{fig:kinematics_lc}
\end{figure*}

Figure~\ref{fig:stacked_spec2D} shows the stacked spectrum, which features several emission lines including the pre--dominantly shock--ionised [OIII, $\lambda$4959], [OIII, $\lambda$5007] and photo--ionised [He II, $\lambda$4686]. The former set of lines are confirmed to originate from the bubble nebula (\citealt{pakull2002}), as noted previously by \citet{roberts2011} and more recently by \citet{zhou2022}, \citet{gurpide2024a} and \citet{gurpide2024b}. We report the velocity shifts and velocity dispersions of all emission lines by computing the wavelength centroid and full width half maxima after fitting them with a Voigt profile for the twelve set of observations: see Table~\ref{tab:gemini_observations} and Figure~\ref{fig:masked_spec1D_ngc1313x2}. Indeed, we find that the [OIII, $\lambda$4959] and [OIII, $\lambda$5007] lines are detected with relatively steady velocities and velocity dispersions, but the [He II, $\lambda$4686] line does show larger variability in both flux and velocity profile. This indicates that the line does originate from regions closer to the ULX (\citealt{grise09}), since its variable flux and radial velocity means that it is strongly ionised by the X--ray flux from the accretion flow, although mass--functions are difficult to obtain since velocity--shifts are not periodic (\citealt{roberts2011}). In order to primarily characterise the continuum of the stacked {\it{Gemini}} spectrum, we masked the emission lines with a Voigt profile $V(x;\sigma)$, estimating the background continuum level from wavelength bins neighbouring the emission lines: 
\begin{multline}
V(\lambda;\sigma) = \int_{-\infty}^{\infty} G(\lambda^{\prime};\sigma) L(\lambda-\lambda^{\prime}; \gamma) d \lambda^{\prime}\\
\end{multline}
where $G(\lambda^{\prime}; \sigma)$ refers to the Gaussian component used to infer gas kinematics and $L(\lambda^{\prime}; \gamma)$ to the Lorentzian used to describe the natural linewidth of the atomic transitions. We subtract emission lines from the optical spectrum to highlight a potentially interesting but marginal feature, corresponding to a sharp decline in flux below $\sim$ 3800 \AA. We suggest that this could be associated with a Balmer break, a known signature of A--type stellar companions. To highlight this clearly, we fit the stacked spectrum with models from stellar libraries such as the Castelli--Kurucz atlas (\citealt{castellikurucz03}). We allow the stellar radius, surface gravity ($\log g = 0.0 - \log g = 50.0$), abundance and surface temperature ($T_{\text{eff}} = 3500 - 45000$ K) to be free parameters of the fit, while fixing the source distance to 3.9 Mpc. The ratio of the H$\alpha$ to H$\beta$ flux (also known as a Balmer decrement which we emphasise is distinct to a Balmer break), and the relation between $N_{\text{H}}$ and $A_{V}$, was used to determine the line--of--sight extinction $E(B-V)$ toward the source of 0.13 $\pm$ 0.03. Here, we assume a Galactic angle--averaged ratio $R_{v} \sim 3.1$; \citealt{cardelli1989}; \citealt{grise09}) and use the above value of $E(B-V)$ to redden the stellar templates. Finally, following \citet{ripamonti11} and \citet{gurpide2024b} we use a sub--solar abundance prior of $\log$ [M/H] $=-2.5$, while considering abundances ranging from $\log$ [M/H] $=-2.5$ to $\log$ [M/H] $=+0.5$ in the fit. Then, we show a sample of five stellar templates that yield the most acceptable reduced $\chi^{2}$ in Figure~\ref{fig:masked_spec1D_ngc1313x2}. The deduced stellar radii range from $R_{*} = 50 - 60 R_{\odot}$ as shown in Figure~\ref{fig:masked_spec1D_ngc1313x2}. We note that the absolute B--band and V--band magnitudes of the optical counterpart are $-4.55$ and $-4.8$ respectively when averaged over all twelve observations, yielding a colour excess $B-V = +0.25$. These parameters suggest a low--mass A--type star, specifically of type A6--A7, showing supergiant--like surface gravity but a luminosity consistent with a bright giant. Importantly, we note that the data is typical of an evolved post--AGB, stripped--envelope star rather than a standard massive super--giant and is thus influenced by binary evolution. It is thus important to consider analytical physics from interacting binary population--synthesis simulations when estimating the formation rates of BH and NS ULXs (see e.g. \citealt{riley2022} and \citealt{misra2023}). We compute the time evolution of the optical luminosity (see Figure~\ref{fig:kinematics_lc}) by integrating in accordance with the equation below:
\begin{align*}
	M_{\text{bol}} - M_{\odot} &= -2.5\log_{10}\bigg(\frac{L_{\text{bol}}}{L_{\odot}} \bigg)\\
	L_{\text{bol}}(t) &= \frac{1}{T}\int_{t_1}^{t_2} L_{\text{bol}}(t) dt
\end{align*}
where the solar luminosity $L_{\odot} = 3.8 \times 10^{33}$ erg s$^{-1}$ is defined as usual. As a sanity check, we re--visit the claim that the ULX hosts an A--type super--giant star, by considering models for the current age of the ULX bubble and compare this to the age of the binary system according to results derived from binary population synthesis. The demographics of X--ray binaries are simulated using the {\tt{posydon}} code, which evolves the binary from zero--age--main--sequence (ZAMS) and then adopts standard physical prescriptions for 1D stellar evolution using non--conservative mass and angular momentum transfer, turbulent mixing in rotationally supported stellar convective envelopes, supernova kicks, common envelope evolution and simple accretion physics for super--critical discs. The results from including these analytical models found that the probability to form a ULX containing a neutron star is much higher for ages $\lesssim 4$ Myr than in the later stages of binary evolution (i.e. $> 20$ Myrs) where black holes begin to dominate. From such a perspective, it is thus not unusual for neutron star pulsating ULXs to contain A--type supergiant stars with considerably smaller ages ranging between $0.5-10$ Myrs. Here we discuss how the sensitivity of the {\it{Gemini}} detector might influence the stellar parameters. We note that the data can support hot OB--type companions if the Balmer break is excluded from the fit by excising wavelengths below 3800 Angstrom. But, A--type stars are predominantly favoured otherwise. Hence, including the Balmer break means that early OB type stars are confidently ruled out, providing an update from the analysis when they were considered as probable candidates using only narrow--band SEDs from the 2015--2016 {\it{HST}} data (\citealt{sathyaprakash2022}). We note that irrespective of evolutionary phase, A--type stars do not contribute significantly to the U--band flux, suggesting additional components would be required to explain the emission at these smaller wavelengths, as noted in previous optical studies (\citealt{grise09}; \citealt{Liu2013}; \citealt{motch2014}). We discuss possible scenarios in section 4. 

Crucially, a break in the optical spectrum could also result due to the degrading instrumental response: for example, see the how the sensitivity curve in Figure~\ref{fig:sens_curve} declines with wavelength. To test whether this might affect our results, we multiplied a smooth reddened power--law spectrum without any breaks by the sensitivity curve, finding that the power--law spectrum does turn over but at a much smaller wavelength (i.e. $\sim$ 3350 \AA) compared to the down--turn seen in the ULX spectrum at 3800 \AA. We therefore argue that the Balmer break is indeed suggestive of an A--type supergiant companion. One can also see this by observing the {\it{Gemini}} optical spectrum of other ULXs such as NGC 5204 X--1, which show a much steeper rising continuum, in comparison to the relatively shallower spectrum of NGC 1313 X--2. 

Finally, we use X--ray fluxes obtained from {\it{Swift}}--XRT observations alongside the {\it{Gemini}}--South optical light--curve to investigate the correlation between the optical flux, the [He II, $\lambda$4686] flux and the full--band X--ray flux. We test whether the optical emission could be explained by considering the reprocessing of X--ray photons from the inner accretion flow. We plot the X--ray/optical/He II fluxes between MJDs $55125$--$55190$ in the left hand panel of Figure~\ref{fig:kinematics_lc}. The He II luminosity appears to peak at around 1.5 $\times 10^{35}$ erg s$^{-1}$ at MJD $55127$, which is also coincident with a peak in the optical continuum luminosity $\sim 3.2\times 10^{37}$ erg s$^{-1}$. Both quantities subsequently decline over a 60 day period. The X--ray luminosity co--varies with the optical emisison in a coherent manner, which enabled us to compute the lag--frequency spectrum to deduce a 2--day propagation delay between the optical emission and the X--ray emission. Such a large time delay implies that the optical envrionment can be more complex than simply containing an evolved binary with an A--type star. We emphasise that our results are useful, but the fact that the [He II, $\lambda 4686$] line is difficult to detect and given the X--ray and optical light--curves are under--sampled, we believe that our lag estimates can only improve in future studies.

\section{Discussion}

In this work, we analysed observations of NGC 1313 X--2 obtained from the Chilean {\it{Gemini--South}} observatory over twelve nights to unveil the stellar type of the companion, primarily by modelling the continuum emission. We considered the entire set of stellar templates from the Castelli--Kurucz atlas, placing a grid over the effective temperature, surface gravity, stellar abundance and radius. We found that A--type supergiant companions tend to be favoured statistically when the {\it{Gemini}} spectrum is modelled in the wavelength range 3600--6500 \AA. This is validated through the best--fit reduced $\chi^{2}$. So far, two other PULXs have well known companions, with NGC 7793 P13 being in orbit around a type B9Ia supergiant (\citealt{motch2014}) and NGC 300 ULX--1 with a red supergiant donor (\citealt{heida2019}). The high level of dust reddening towards the edge--on galaxy harbouring the PULX NGC 5907 X--1 precludes the identification of an optical counterpart (\citealt{heida2019a}), and similarly for M82 X--2. We note that the well studied Galactic super--Eddington source SS433 is also believed to contain an evolved A7Ib companion, observed during an eclipse of the optical continuum (\citealt*{gies02} and \citealt{hillwig04}). Such stars could in principle reach sufficiently high mass transfer rates to power the super--Edddington luminosities of ULXs. We have compared the stacked {\it{Gemini--South}} spectra of three distinct ULXs to ascertain whether the observed cut--off in the optical spectrum of NGC 1313 X--2 below 3800 \AA is an instrumental artefact or associated with a true Balmer break. Firstly, we note that the optical spectrum of NGC 5204 X--1 is steeper, and suggests that the sensitivity of the {\it{Gemini--South}} detector is sufficient to be able to measure the decline in optical flux below 3800 \AA by simple comparison. Moreover, the absolute magnitude of the counterpart in the $V$ and $B$ bands are $-4.8$ and $-4.55$ respectively, indicating a colour excess $B-V = +0.25$. This is as expected for an evolved A--type super--giant.

Prior to the detection of pulsations in this ULX, historical studies have proposed an alternative hypothesis to explain its formation history. NGC 1313 is known to be a starburst galaxy (\citealt{moorwood_1996}) and is thought to have interacted with a neighbouring satellite sometime in the recent past (\citealt{okamoto2005}). However, there is no direct evidence for such a scenario due to the lack of an observable companion. However, starburst events leave distinct signatures in the dynamical and chemical evolution of the galaxy. For instance, a past collision should kinematically decouple different regions of the galactic disc such that its azimuthal angular momentum profile varies with radius indicating clear influence from starburst episodes. Such events could also enhance the production of metals from inflated core--collapse supernovae, which is a useful indicator of young high mass X--ray binaries, dynamically evolved stellar clusters and Wolf--Rayet populations and such environments usually have super--solar abundances. While the galaxy NGC 1313 does not show radial metallicity gradients, which is in a similar vein to the Large Magellanic Cloud (\href{https://www.aanda.org/articles/aa/pdf/forth/aa58122-25.pdf}{Ojeda-Diaz, B. et al. 2026}), it does show evidence for a young star cluster and so the ULXs in this galaxy could have formed due to a starburst in the past: $> 20$ Myrs ago. The Very Large Telescope Array {\it{VLT--MUSE}} observations near regions closer to the ULX will provide further ideas on the chemical properties and the influence of starburst events on star--formation.

However, if the ULX did not originally form in its current location, this may support the scenario that both the source and the accompanying star--cluster have been ejected from the galaxy due to gravitational--wave recoil from the central supermassive BH or from a population of stellar mass black holes that have accumulated near the nucleus due to mass segregation. Alternatively, a galaxy like NGC 1313 undergoing starburst episodes may have experienced a super--wind episode (\citealt{lopezrodriguez2021}) that could also form stars through gas compression. This process recycles the stars back into the galaxy, making it more challenging to spatially resolve very old stars from those that recently formed via recycled gas. However, it is likely that the ULX may have formed through in--situ star--formation via the accretion of gas from the intergalactic medium (IGM), and hence represents a standard {\it{extreme}} HMXB given the discovery of coherent pulsations. Moreover, the ULX does not directly show evidence of proper motion since there are no detections of bow--shocks in the optical images at present, albeit they are likely not detectable due to low surface brightness at those distances. Moreover, the ULX is $\lesssim 1$ Myrs old according to the age of its bubble, which, given the vast distances from the plane of the host galaxy (i.e. $\sim 0.01$ Mpc) would rule out gravitational recoil from simple light--travel time arguments (i.e. $t_{\text{eject}} = \frac{d}{c} \gg t_{\text{Hubble}}$ years).

\subsection{Orbital parameters}

\begin{figure}
\includegraphics[scale=0.35]{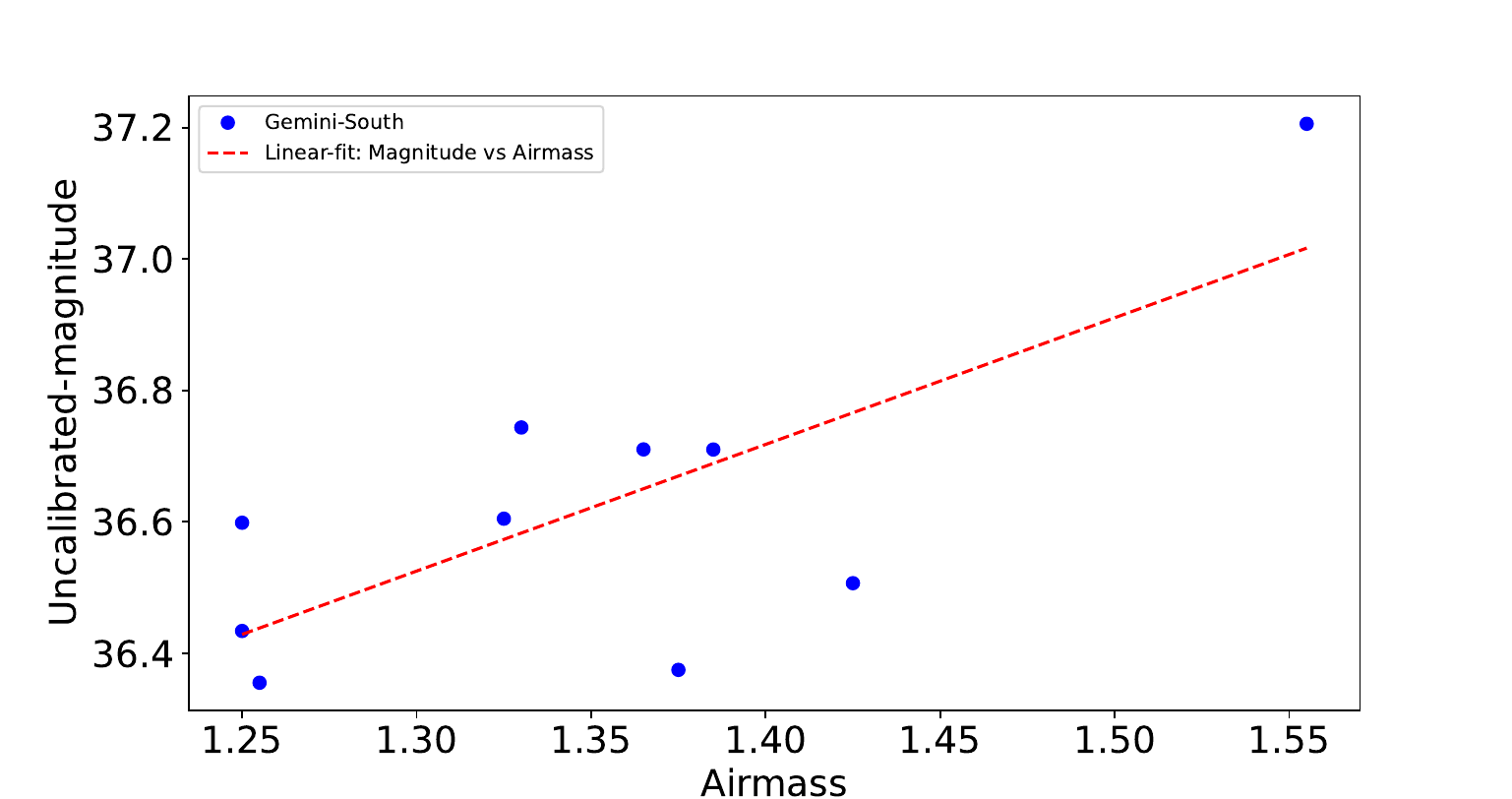}
\caption{Here, we show the optical B--band continuum magnitude as a function of airmass, with the dashed red--line indicating the best fit line, which was used to derive the airmass k--correction $k= 2.7 \pm 0.6$.}
\label{fig:airmag_relation}
\end{figure}

The inferred stellar radius and the fact that the primary is known to be a neutron star (with $M_1 = 1.4$ $M_{\odot}$) implies that we can constrain the range of possible orbital separations, eccentricities and companion masses if we assume that the companion is filling its Roche Lobe. The expression for the Roche Lobe radius $R_{L}$ as a function of binary mass ratio $q = M_2/M_1$, eccentricity $e$, true anomaly $\nu$ and orbital separation $a$ (\citealt{eggleton1983}) is:

\begin{equation}
\frac{R_L}{a} = \bigg[ \frac{(1 - e^{2})}{1 + e\cos \nu} \bigg ]\bigg[ \frac{0.49q^{2/3}}{0.6q^{2/3} + \ln (1 + q^{1/3} )} \bigg]
\end{equation}

We plot the range of possible primary masses $M_2$ and orbital separations $a$ (see Figure~\ref{fig:orb_constraints}) assuming that the stellar radius should be between 0.9--1.1 times the Roche Lobe radius in order to initiate significant mass transfer: this allows us to exclude orbital solutions if the stellar radius is below 0.9$R_{L}$ and above 1.1$R_{L}$. This range is somewhat arbitrary, but the intention is to require the stellar radius to be as close to $R_{L}$ without the star under--filling or over--filling the Roche Lobe. We emphasise that the orbital separation $a$ defined here is distinct from the projected orbital separation $a_p = a\sin i$ usually inferred from the amplitude of radial velocity curves. So, if the latter is known, the inclination to the system could in principle be estimated. However, as shown in Figure 6 (left panel) there do not appear to be any periodicities associated with the [He II, $\lambda$ 4686] emission line.  

We consider two extreme cases for the Roche Lobe radius when the donor is at the periastron and apastron, with the range in stellar radius being 45 - 70 $R_{\odot}$, while the precise stellar radii we obtain from fitting the {\it{Gemini}} spectrum are shown in Table~\ref{Tab:stellar_fits}. Finally, we caution that the shape of the Roche Lobe can also be influenced by the donor star's rotation (\citealt{sepinsky07}), which we currently neglect. Under these assumptions, the range of allowed orbital separations, donor masses and eccentricities are shown in Figure~\ref{fig:orb_constraints}. Irrespective of companion mass, orbital phase or eccentricity, we exclude orbital separations $< 50 R_{\odot}$ and $> 200 R_{\odot}$.

\subsection{Blue excess}

The {\it{HST}} photometric data reveal an excess in emission blueward of 3500 Angstrom with respect to the models considered above, which requires an additional spectral component if the {\it{Gemini}} data are well-fit by an A-type star. This could plausibly be explained by emission from an irradiated accretion disc, or reprocessing of X--ray photons from the donor star itself (given the possible interpretation of its optical magnitude variability as arising due to ellipsoidal modulations; \citealt{zampieri2012}) or the extended photosphere of an optically thick wind. It has also been noted that the B-band optical emission varies stochastically over weekly timescales, requring a non-stellar origin (\citealt{grise09}). \cite{sathyaprakash2022} attempted to constrain the reprocessed fraction $f_{\text{irr}}$ by modelling the UV/optical/NIR SED with irradiated disc models, but found $f_{\text{irr}}$ to be unreasonably large if one assumes that the X--ray photons are reprocessed from a standard thin disc, and an unrealistically large $\log \dot{m} > 4.0$ for a radiatively inefficient slim disc, albeit without reprocessing. Moreover, \citet{qiufeng2021} noted that the SED of NGC 1313 X--2 can be reasonably well modelled if one considers the scattering of hard X--rays emitted from the inner regions of a super-Eddington disc in addition to emission from the wind photosphere. 3D simulations of super-Eddington discs have been applied to model the optical emission from tidal disruption events using the {\it{Athena}++} code (\citealt{athena_code_dev}), and such codes could in principle be used in future studies to properly explain the contribution from disc reprocessing in the accretion flows of ULXs. At infrared wavelengths, the {\it{HST}} emission also shows an excess with respect to the stellar continuum and was found to be variable (within a three month period), hinting additional contributions from circumbinary dust surrounding the ULX or synchrotron emission from a jet, among other possible scenarios discussed in \citet{lau2019}. 

\subsection{X--ray and optical variability}

\begin{figure*}
\includegraphics[scale=0.5]{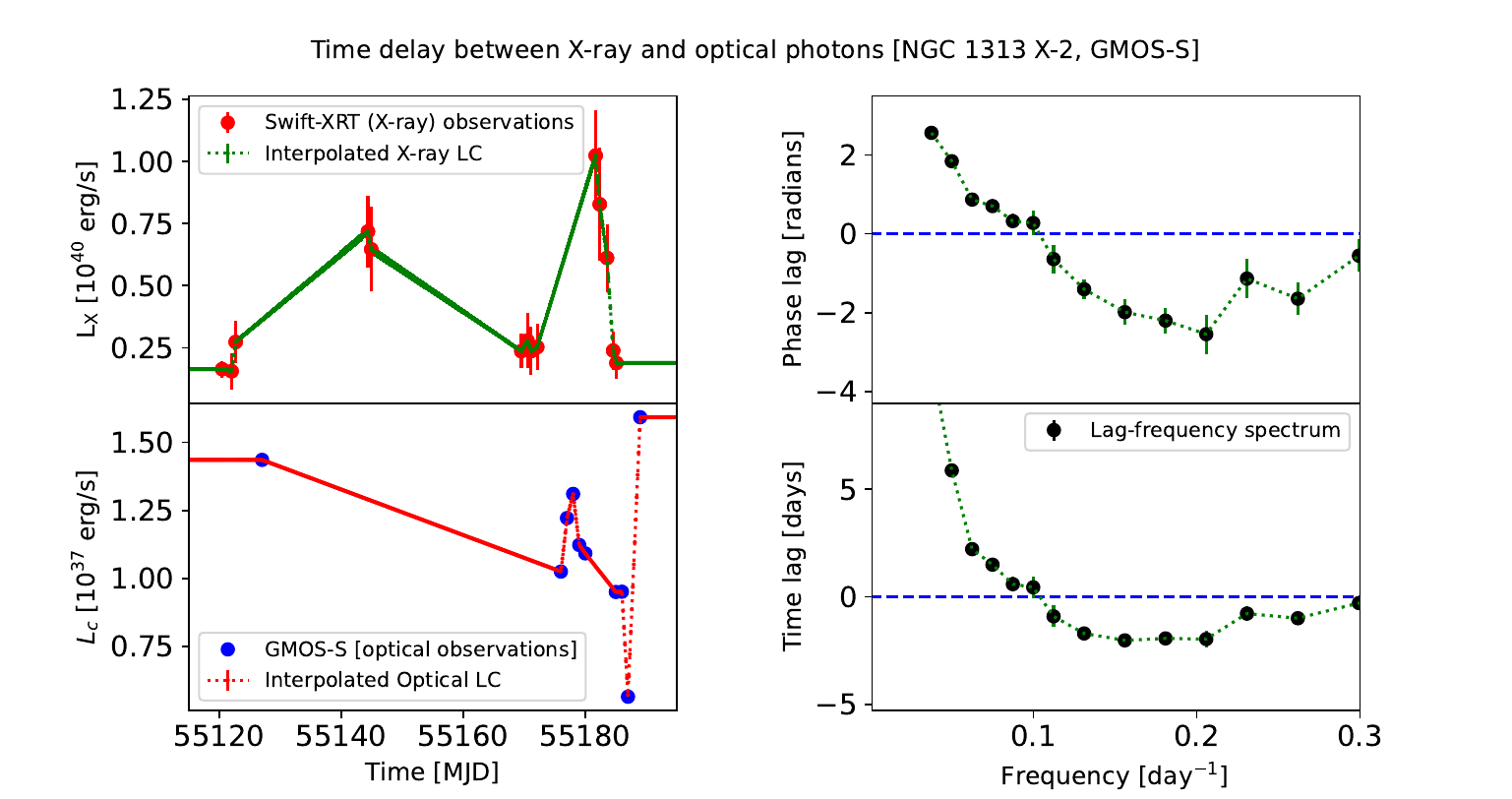}
\caption{{\it{Left}}: Interpolated X--ray and optical light--curves from {\it{Swift}--XRT} (top) and {\it{Gemini--South}} (bottom). {\it{Right}}: Lag--frequency spectra showing the phase delay (top) and time delay (bottom) between the X--ray and optical continua as a function of frequency, ensuring to correct for phase--wrapping. The lag--frequency spectrum is structured, but observational gaps in the optical continuum means that the light--curve is not perfectly sampled. However, we are confident of observing a 2--day propagation lag between the X--ray and optical emission, at--least within the timespan of the observations shown here, which indicates a complex optical environment. We emphasise that the correlation between the X--ray and optical emission implies that the optical flux is likely un--affected by changes in airmass.}
\label{fig:lag_freq_spec}
\end{figure*}

The size scale of the accretion flow is constrained by an apparent 2 day delay between the {\it{Gemini--South}} (optical) and {\it{Swift}}--XRT (X--ray) luminosity, which is in line with the methods discussed in \cite{uttley_2014}. We accounted for the non--uniform sampling of the optical continuum by interpolating the {\it{Gemini}} optical data with a cubic spline. As shown in Figure~\ref{fig:lag_freq_spec}, the lag between the optical and X--ray emission appears clear in the lag--frequency spectrum, suggesting a propagation delay of $\lesssim 2$ days. However, the optical and X--ray fluxes are under--sampled and may not be consistent with the true noise process of the data. Hence, it is difficult to make precise statements about disc sizes using {\it{Gemini}} and {\it{Swift}--XRT} observations. If one believes the observed $\sim 2$ day lag between the X--ray and optical data, and the disc size is computed using light--travel time arguments (i.e. $R_{d} = c\tau$), this implies that $R_{d}$ should be $\lesssim 5 \times 10^{13}$ meters or $\lesssim 1 \times 10^{10}$ gravitational radii ($R_{g}$), if one assumes a 1.5 $M_{\odot}$ NS. Such large delays or disc sizes have been observed in the rising hard state of typical HMXBs or Be X--ray binaries (\citealt{dudizan2025}), in which the hard X--ray and radio emission lags between the optical flux. This phenomenon is usually explained in terms of the disc instability model. However, such mechanisms may not be valid for super--Eddington discs, other viscous dissipation mechanisms may apply, additional emission components may be required or the compact object may not necessarily be an NS. This would then indicate that the reprocessing region implied by our analysis could originate from a much larger circumbinary disc, jets, or correspond to the surface of a stellar companion in a wider orbit than implied by the Roche--Lobe geometry argument.  

Finally, we test whether the variability in the optical B--band luminosity could systematically be driven by changes in airmass, which is a measure of atmospheric extinction, rather than some underlying physical mechanism. We computed median airmass values for each observing night and then plotted these airmass values against the uncalibrated B--band magnitude, as shown in Figure~\ref{fig:airmag_relation}. The data points indicate a moderately strong correlation between airmass and observed B--band continuum magnitude. We perform a simple linear fit, relating the airmass $X$ and magnitude $m_{\text{X}}$, where: 
\begin{equation}
m_{\text{X}} (X) = kX + m_{0}
\end{equation}
to obtain a Spearman--rank coefficient $\rho = 0.66$ ($p$--value $= 0.04$) and a Pearson coefficient of $r = 0.85$ ($p$--value $= 0.002$). The atmospheric extinction coefficient $k$ is the gradient of the best--fit linear model, where in this case, we obtain $k= 1.9 \pm 0.6$ uncalibrated mag airmass$^{-1}$. Hence, we can determine the amount by which we expect the optical flux to maximally degrade due to atmospheric extinction by 
\begin{equation}
R = 10^{-0.4kX} 
\end{equation}
with $R = F/F_{0}$ being the ratio between the minimum and maximum B--band flux. This suggests a $\sim 60$ \% variation in optical flux due to airmass alone, whereas the observed maximal flux variability is $\sim 400$\%. Moreover, since the optical flux is correlated with the X--ray emission, we consider it unlikely that airmass is the only parameter that modulates the optical emission. We can perform a similar experiment as a function of dust--reddening $A_{v}$, relating this to the column density $N_{\text{H}}$, although here we are limited by the non--simultaniety of the X--ray and optical data and rely on interpolation schemes to infer the gas column density. We note that the observed variability is non--radial and we could attribute the maximal velocity amplitude to changes in airmass or dust--reddening, although to obtain a more complete picture we need better sampling of the optical light--curve. This can be achieved with future {\it{VLT}} or {\it{LSST}} observations given their sufficient sensitivities in the U,V,B,R,I bands. These future missions would allow us to test for ellipsoidal modulation or disc precession mechanisms more precisely and thereby constrain formation scenarios.

\subsection{Emission line kinematics}

\begin{figure*}
\centering
\includegraphics[scale=0.5]{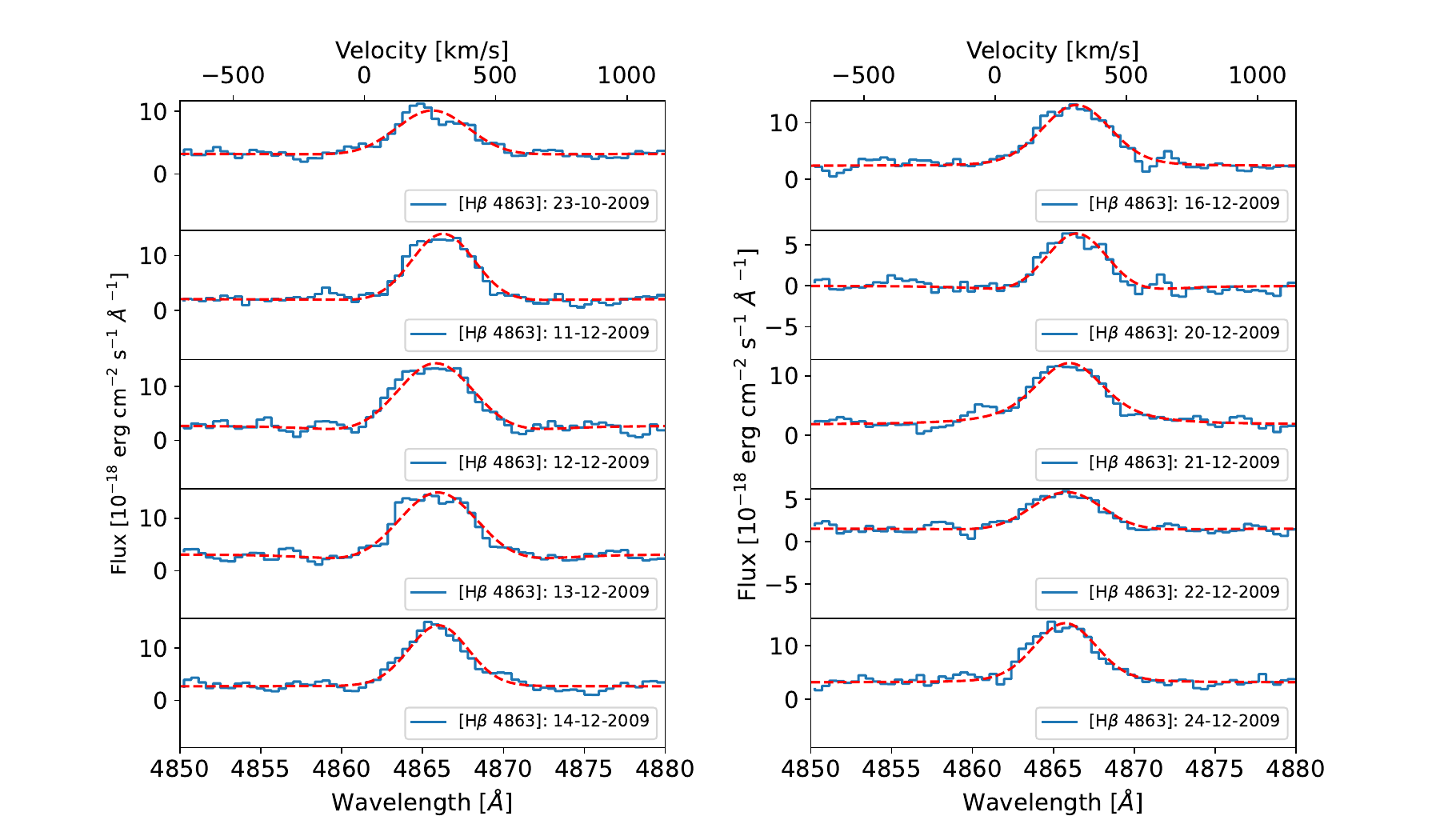}
\caption{Here, we show Voigt profiles (dashed red lines) fit to the nebular [H$\beta$, $\lambda$4863] emission lines (blue steps) for all ten observations conducted in 2009, using least squares minimisation. This allowed us to deduce the radial velocity of the nebula after subtracting the systemic velocity of the host galaxy (470 km s$^{-1}$) from the velocity of the H$\beta$ line via the non--relativistic Doppler shift equation. The line profile appears steady in all observations, displaying a constant velocity shift around 180 km s$^{-1}$, indicates steady velocity dispersions, meaning that the bubble expansion is adiabatic relative to dynamical motions associated with the ULX counterpart, as suggested by the variability in the [HeII, $\lambda$4686] line (see below). We compute the line flux via numerical integration.}
\label{fig:nebular_lines1}
\end{figure*}

\begin{figure*}
\centering
\includegraphics[scale=0.5]{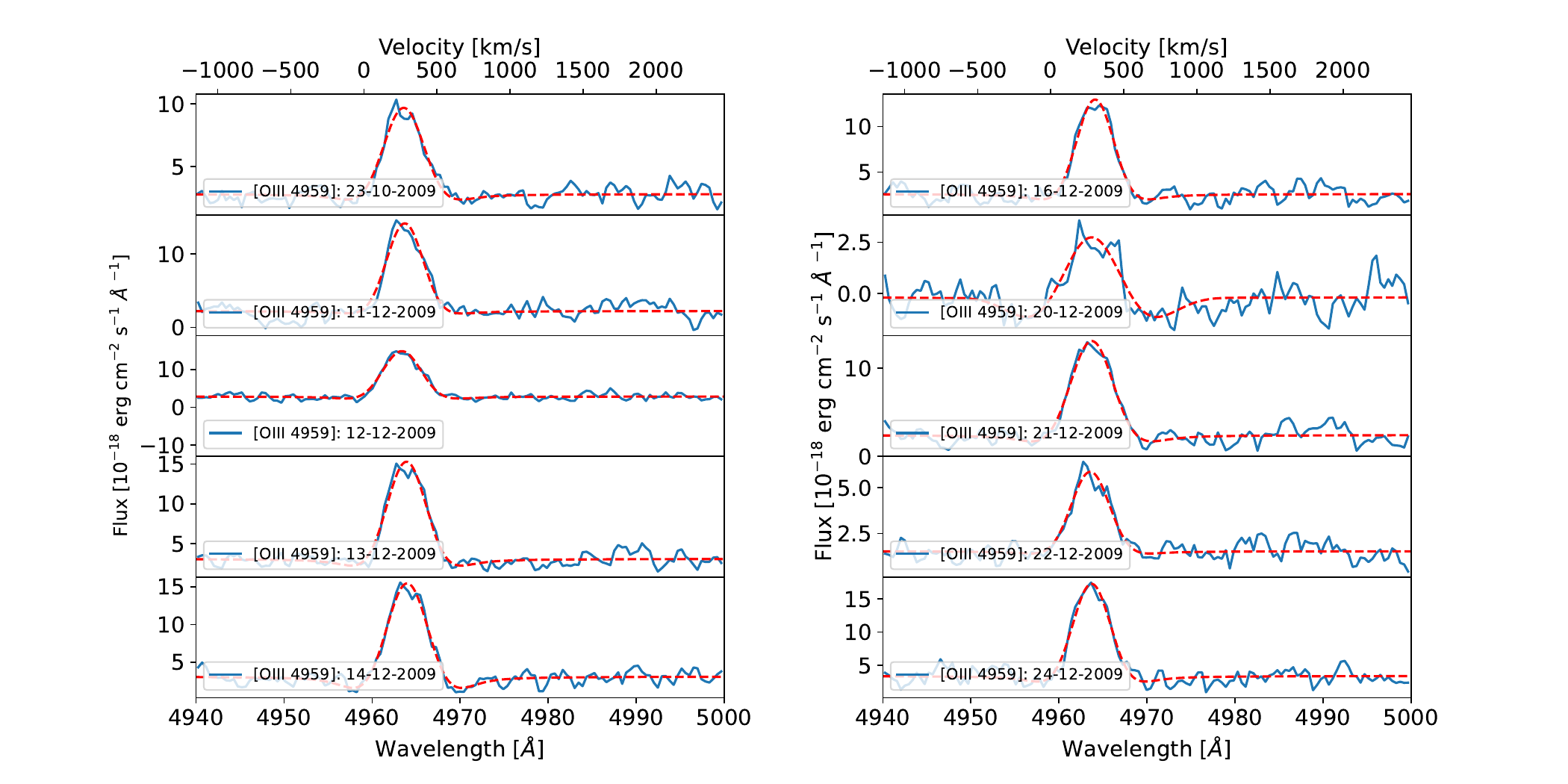}
\caption{Same as Figure~\ref{fig:nebular_lines1} but for the [OIII, $\lambda$4959] line.}
\label{fig:nebular_lines2}
\end{figure*}

\begin{figure*}
\centering
\includegraphics[scale=0.5]{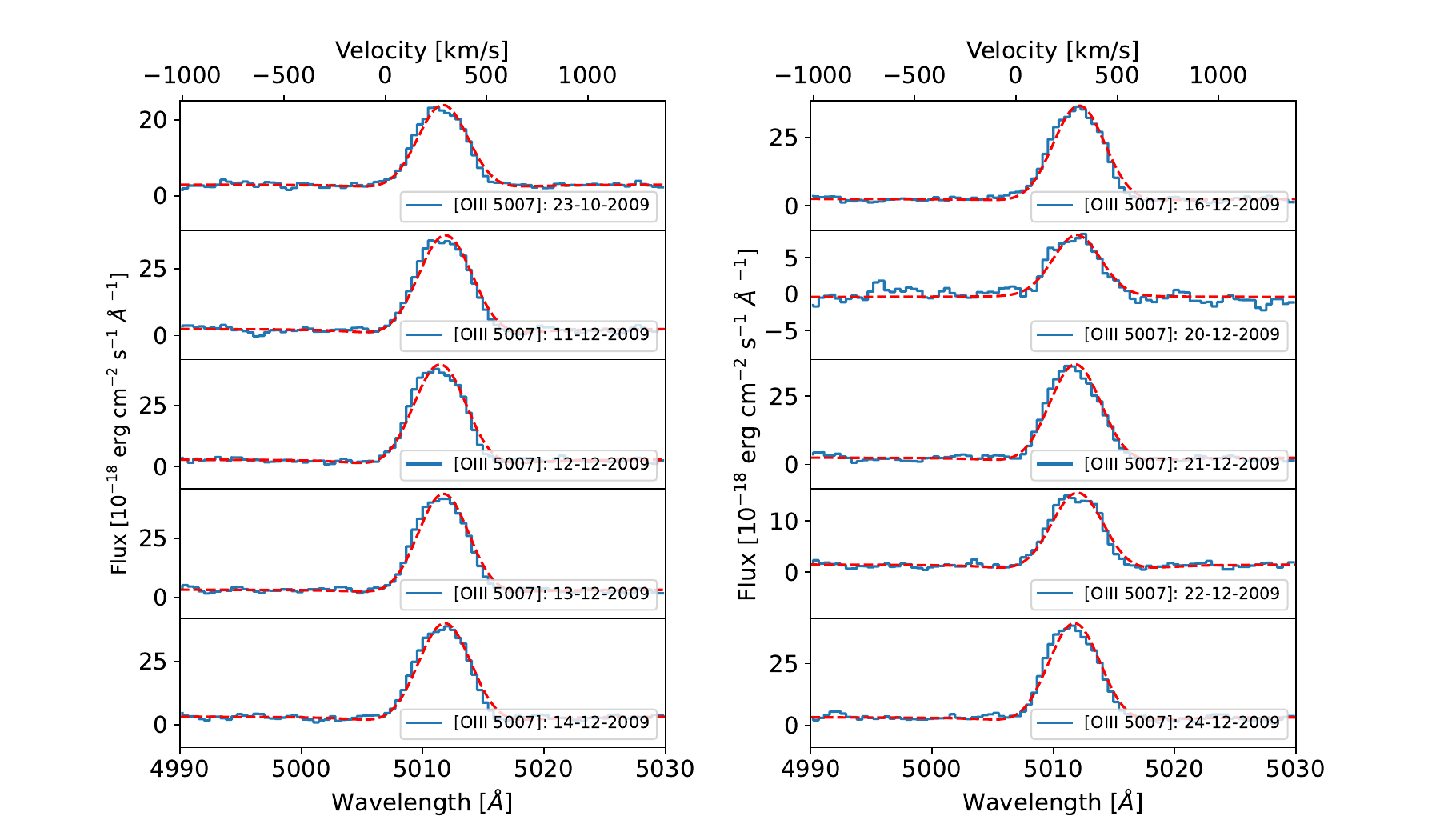}
\caption{Same as Figure~\ref{fig:nebular_lines1} but for the [OIII, $\lambda$5007] line.}
\label{fig:nebular_lines2}
\end{figure*}
\begin{figure*}
\centering	
\includegraphics[scale=0.5]{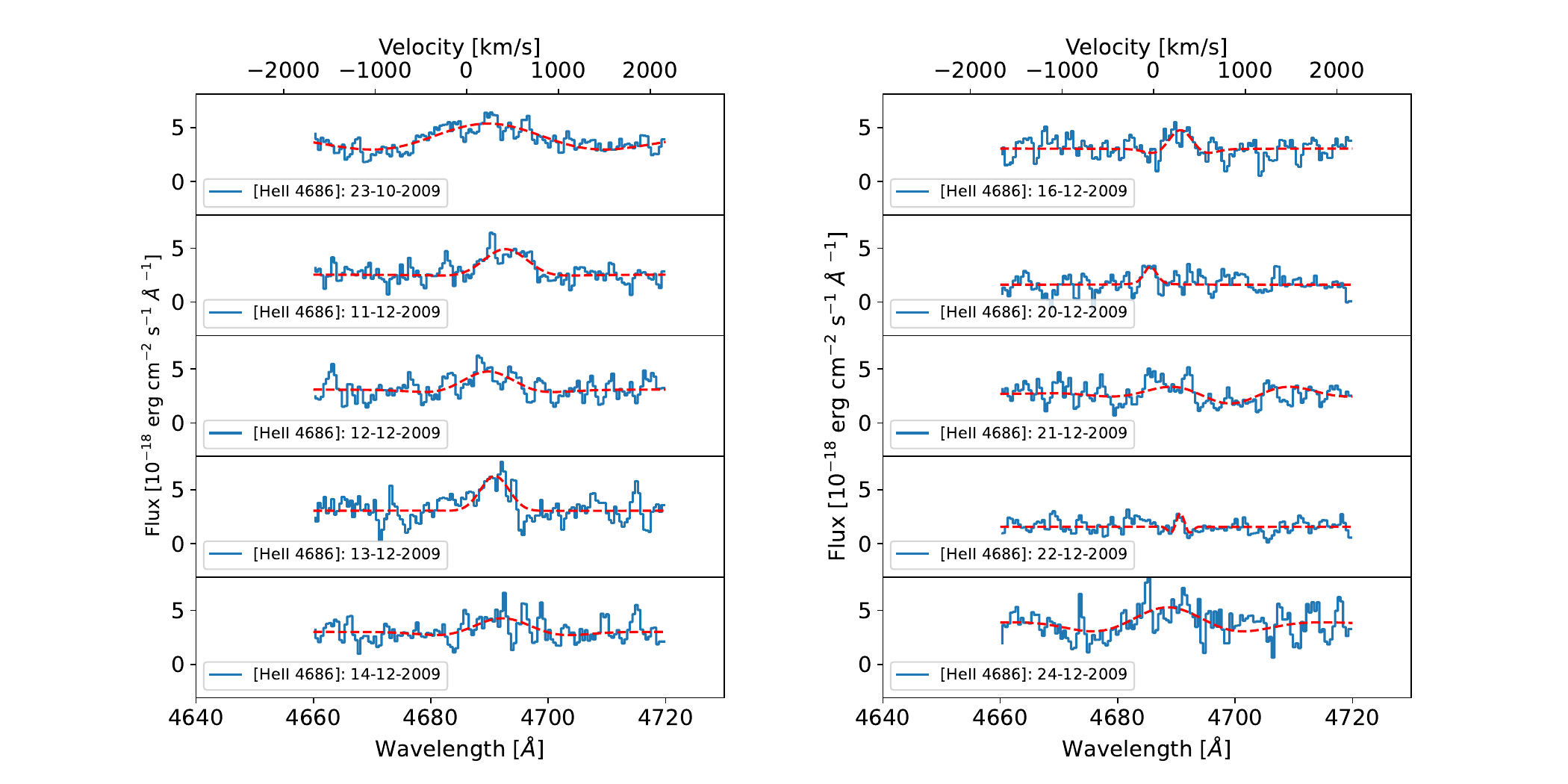}
\caption{Same as Figure~\ref{fig:nebular_lines1} but for the [He II, $\lambda$4686] emission line associated with the ULX counterpart. The emission line is faint in these observations and difficult to characterise. Generally, the line appears broad on the first observation night, then steadily decreases in luminosity, displaying non--periodic scatter in radial velocity and sometimes features a doublet, as expected for a Keplerian thin disc. However, one must treat these variations with caution owing to the faintness of the line, even though the calibrated spectrum is accurate.}
\label{fig:nebular_lines3}
\end{figure*}

Here, we begin to test for the above scenarios by investigating how the emission line profile varies across these 10--12 observations. After subtracting the systemic velocity of the host galaxy, which we assume to be $\sim 470 \pm 2$ km/s (\citealt{seheon2018}), the line--of--sight velocities and velocity dispersions for the three characteristic emission lines ([OIII] $\lambda$5007, [OIII] $\lambda$4959 and [He II] $\lambda$4686) are shown in Figure~\ref{fig:kinematics_lc} and the associated line profiles in Figures~\ref{fig:nebular_lines2} and~\ref{fig:nebular_lines3}. The [OIII] emission lines arise from more extended regions surrounding the ULX and are associated with shock ionised nebulae, although a narrow emission line core indicates that some fraction of the line arises due to photo--ionisation (\citealt{zhou2022}; \citealt{soria2006}; \citealt{pakull2008}; \citealt{pakull2006}). However, since the [He II, $\lambda$4686] line is more variable, it is thought to be directly associated with the ULX. In order to constrain the orbital period from both the broad--band optical continuum (3500--6000 Angstrom) and the He II emission line centroid, we applied an accelerated search for optical pulsations using {\tt{presto}}. However no periodicities were found in the optical emission, which is in line with earlier studies from \cite{roberts2011}. We estimate the detection significance to be $< 2\sigma$ after considering the total number of trials in the PSD. We believe that the inferred stellar radii, as discussed above, provide more useful information to this study. The He II emission line appears to feature a doublet at some epochs, which could indicate disc motions, and we aim to provide more detailed information in future studies using a reverberation mapping code that use velocity--delay maps to resolve the disc structure. Such techniques have been successfully demonstrated in mapping the broad line regions of active galactic nuclei (e.g. \citealt{horne2021}; \citealt{ingram2016}). 
\newline
\newline
The age of the bubble surrounding NGC 1313 X--2 is indicative of recent star formation in the outskirts of the galaxy, aside from regular, old and metal--poor stars. The bubble was thought to have been excavated by radiative and mechanical feedback from the ULX alone, possibly polluted by the nearby stellar population. The age of the bubble scales with the radiative flux of the H$\beta$ line, and can be estimated using the \cite{weaver1977} prescription:
\begin{equation}
t = \frac{3}{5} R_{\text{b}} \bigg ( \frac{1}{100 \text{km/s}} \bigg ) \bigg [ \bigg ( \frac{S_{\text{H $\beta$}}}{\text{erg cm$^{-2}$ s$^{-1}$}} \bigg ) \bigg ( \frac{1}{7.44 \times 10^{-6}} \bigg ) \bigg ( \frac{\text{cm}^3} {n_{\text{ism}}} \bigg ) \bigg ]^{2.41}
\end{equation}
In line with previous studies, such an age is consistent with the expected value for a ULX undergoing a super--Eddington mass transfer phase: i.e. $\sim 1$ Myr, assuming a scenario where it is associated with a nearby stellar population.

\section{Conclusions}

In summary, the paper provides constraints on the type of the stellar companion to the pulsating ULX NGC 1313 X--2, by considering twelve nights of {\it{Gemini}} data. Upon stacking the spectra from all available nights using average $\sigma$--clipping methods and masking cosmic--rays, we believe that the break in the spectrum below 3800 Angstrom could be explained by an evolved A--type supergiant by fitting the optical continuum with models from the Castelli--Kurucz stellar atlas. We include a grid that comprises surface gravities, stellar radii, metal abundances and surface temperatures of the star. In such a scenario, we are able to provide some approximate constraints on the orbital parameters of the ULX binary, where limits on the orbital separation, companion mass and eccentricity can be derived, assuming that the star is filling its Roche--Lobe. We believe the U--band flux might be better explain if the SED models also consider emission from outer disc winds, circumbinary discs, jets and/or the irradiated face of the secondary star. However, these uncertainties remain and in this particular case we find some evidence for these additional scenarios while ruling out others given a 2--day propagation lag between the X--ray and optical emission, the absence of strict periodic variations in the optical light--curve and the velocity of the He II line. We note that the formation history of the ULX may be influenced by super--wind episodes and disturbed galaxy kinematics in its location in the outskirts of the galaxy. However, we suggest that it likely belongs to a standard {\it{extreme}} high mass X--ray binary given the observed X--ray pulsations and the lack of a cut--off in the X--ray luminosity function according to previous {\it{NuSTAR}} results.       

\section{Acknowledgements}
R.S. acknowledges financial support from the Italian Ministry for University and Research, through the grants 2022Y2T94C (SEAWIND) and from INAF through LG 2023 BLOSSOM. T.P.R. acknowledges support from the Science and Technology Facilities Council (STFC) as part of the consolidated grant award ST/X001075/1. 

\section{Data Availability}
The data used in this manuscript were downloaded from the {\it{Gemini}} Observatory Archive\footnote{https://archive.gemini.edu/searchform}, the MAST portal\footnote{https://mast.stsci.edu/portal/Mashup/Clients/Mast/Portal.html} and the online {\it{Swift}}--XRT lightcurve builder (\citealt{evans2007}). All data can be made available upon reasonable request to the authors.

\bibliographystyle{mnras}
\bibliography{biblio}

@ARTICLE{kingmiddleton23,
       author = {{King}, Andrew and {Lasota}, Jean-Pierre and {Middleton}, Matthew},
        title = "{Ultraluminous X-ray sources}",
      journal = {\nar},
         year = 2023,
        month = jun,
       volume = {96},
          eid = {101672},
        pages = {101672},
          doi = {10.1016/j.newar.2022.101672},
archivePrefix = {arXiv},
       eprint = {2302.10605},
 primaryClass = {astro-ph.HE},
       adsurl = {https://ui.adsabs.harvard.edu/abs/2023NewAR..9601672K}
}

@ARTICLE{dudizan2025,
       author = {{Du}, Dizhan and {You}, Bei and {Yan}, Zhen and {Cao}, Xinwu and {Hameury}, Jean-Marie and {Wu}, Yue},
        title = "{A comprehensive study of time delay between optical/near-infrared and X-ray emissions in black hole X-ray binaries}",
      journal = {\mnras},
         year = 2025,
        month = nov,
       volume = {543},
       number = {3},
        pages = {2575-2598},
          doi = {10.1093/mnras/staf1607},
archivePrefix = {arXiv},
       eprint = {2507.00578},
 primaryClass = {astro-ph.HE},
       adsurl = {https://ui.adsabs.harvard.edu/abs/2025MNRAS.543.2575D}
}

@software{pyraf2012,
       author = {{Science Software Branch at STScI}},
        title = "{PyRAF: Python alternative for IRAF}",
 howpublished = {Astrophysics Source Code Library, record ascl:1207.011},
         year = 2012,
        month = jul,
          eid = {ascl:1207.011},
archivePrefix = {ascl},
       eprint = {1207.011},
       adsurl = {https://ui.adsabs.harvard.edu/abs/2012ascl.soft07011S}
}

@ARTICLE{riley2022,
       author = {{Riley}, Jeff and {Agrawal}, Poojan and {Barrett}, Jim W. and {Boyett}, Kristan N.~K. and {Broekgaarden}, Floor S. and {Chattopadhyay}, Debatri and {Gaebel}, Sebastian M. and {Gittins}, Fabian and {Hirai}, Ryosuke and {Howitt}, George and {Justham}, Stephen and {Khandelwal}, Lokesh and {Kummer}, Floris and {Lau}, Mike Y.~M. and {Mandel}, Ilya and {de Mink}, Selma E. and {Neijssel}, Coenraad and {Riley}, Tim and {van Son}, Lieke and {Stevenson}, Simon and {Vigna-G{\'o}mez}, Alejandro and {Vinciguerra}, Serena and {Wagg}, Tom and {Willcox}, Reinhold and {Team Compas}},
        title = "{Rapid Stellar and Binary Population Synthesis with COMPAS}",
      journal = {\apjs},
         year = 2022,
        month = feb,
       volume = {258},
       number = {2},
          eid = {34},
        pages = {34},
          doi = {10.3847/1538-4365/ac416c},
archivePrefix = {arXiv},
       eprint = {2109.10352},
 primaryClass = {astro-ph.IM},
       adsurl = {https://ui.adsabs.harvard.edu/abs/2022ApJS..258...34R}
}

@ARTICLE{ingram2016,
       author = {{Ingram}, Adam and {van der Klis}, Michiel and {Middleton}, Matthew and {Done}, Chris and {Altamirano}, Diego and {Heil}, Lucy and {Uttley}, Phil and {Axelsson}, Magnus},
        title = "{A quasi-periodic modulation of the iron line centroid energy in the black hole binary H1743-322}",
      journal = {\mnras},
         year = 2016,
        month = sep,
       volume = {461},
       number = {2},
        pages = {1967-1980},
          doi = {10.1093/mnras/stw1245},
archivePrefix = {arXiv},
       eprint = {1607.02866},
 primaryClass = {astro-ph.HE},
       adsurl = {https://ui.adsabs.harvard.edu/abs/2016MNRAS.461.1967I}
}

@ARTICLE{okamoto2005,
       author = {{Okamoto}, Takashi and {Eke}, Vincent R. and {Frenk}, Carlos S. and {Jenkins}, Adrian},
        title = "{Effects of feedback on the morphology of galaxy discs}",
      journal = {\mnras},
         year = 2005,
        month = nov,
       volume = {363},
       number = {4},
        pages = {1299-1314},
          doi = {10.1111/j.1365-2966.2005.09525.x},
archivePrefix = {arXiv},
       eprint = {astro-ph/0503676},
 primaryClass = {astro-ph},
       adsurl = {https://ui.adsabs.harvard.edu/abs/2005MNRAS.363.1299O}
}

@ARTICLE{lopezrodriguez2021,
       author = {{Lopez-Rodriguez}, Enrique and {Guerra}, Jordan A. and {Asgari-Targhi}, Mahboubeh and {Schmelz}, Joan T.},
        title = "{The Strength and Structure of the Magnetic Field in the Galactic Outflow of Messier 82}",
      journal = {\apj},
         year = 2021,
        month = jun,
       volume = {914},
       number = {1},
          eid = {24},
        pages = {24},
          doi = {10.3847/1538-4357/abf934},
archivePrefix = {arXiv},
       eprint = {2102.03362},
 primaryClass = {astro-ph.GA},
       adsurl = {https://ui.adsabs.harvard.edu/abs/2021ApJ...914...24L}
}

@ARTICLE{pintowalton23,
       author = {{Pinto}, Ciro and {Walton}, Dominic J.},
        title = "{Ultra-luminous X-ray sources: extreme accretion and feedback}",
      journal = {arXiv e-prints},
         year = 2023,
        month = jan,
          eid = {arXiv:2302.00006},
        pages = {arXiv:2302.00006},
          doi = {10.48550/arXiv.2302.00006},
archivePrefix = {arXiv},
       eprint = {2302.00006},
 primaryClass = {astro-ph.HE},
       adsurl = {https://ui.adsabs.harvard.edu/abs/2023arXiv230200006P}
}

@ARTICLE{pintokosec23,
       author = {{Pinto}, C. and {Kosec}, P.},
        title = "{Winds in ultraluminous X‑ray sources: New challenges}",
      journal = {Astronomische Nachrichten},
         year = 2023,
        month = may,
       volume = {344},
       number = {4},
          eid = {e20220134},
        pages = {e20220134},
          doi = {10.1002/asna.20220134},
archivePrefix = {arXiv},
       eprint = {2211.00014},
 primaryClass = {astro-ph.HE},
       adsurl = {https://ui.adsabs.harvard.edu/abs/2023AN....34420134P}
}

@ARTICLE{earnshaw2019,
       author = {{Earnshaw}, H.~P. and {Roberts}, T.~P. and {Middleton}, M.~J. and {Walton}, D.~J. and {Mateos}, S.},
        title = "{A new, clean catalogue of extragalactic non-nuclear X-ray sources in nearby galaxies}",
      journal = {\mnras},
         year = 2019,
        month = mar,
       volume = {483},
       number = {4},
        pages = {5554-5573},
          doi = {10.1093/mnras/sty3403},
archivePrefix = {arXiv},
       eprint = {1812.04684},
 primaryClass = {astro-ph.HE},
       adsurl = {https://ui.adsabs.harvard.edu/abs/2019MNRAS.483.5554E}
}

@ARTICLE{kovlakas2020,
       author = {{Kovlakas}, K. and {Zezas}, A. and {Andrews}, J.~J. and {Basu-Zych}, A. and {Fragos}, T. and {Hornschemeier}, A. and {Lehmer}, B. and {Ptak}, A.},
        title = "{A census of ultraluminous X-ray sources in the local Universe}",
      journal = {\mnras},
         year = 2020,
        month = nov,
       volume = {498},
       number = {4},
        pages = {4790-4810},
          doi = {10.1093/mnras/staa2481},
archivePrefix = {arXiv},
       eprint = {2008.10572},
 primaryClass = {astro-ph.GA},
       adsurl = {https://ui.adsabs.harvard.edu/abs/2020MNRAS.498.4790K}
}

@ARTICLE{walton2022,
       author = {{Walton}, D.~J. and {Mackenzie}, A.~D.~A. and {Gully}, H. and {Patel}, N.~R. and {Roberts}, T.~P. and {Earnshaw}, H.~P. and {Mateos}, S.},
        title = "{A multimission catalogue of ultraluminous X-ray source candidates}",
      journal = {\mnras},
         year = 2022,
        month = jan,
       volume = {509},
       number = {2},
        pages = {1587-1604},
          doi = {10.1093/mnras/stab3001},
archivePrefix = {arXiv},
       eprint = {2110.07625},
 primaryClass = {astro-ph.HE},
       adsurl = {https://ui.adsabs.harvard.edu/abs/2022MNRAS.509.1587W}
}

@ARTICLE{fabbiano2001,
       author = {{Fabbiano}, G. and {Zezas}, A. and {Murray}, S.~S.},
        title = "{Chandra Observations of ``The Antennae'' Galaxies (NGC 4038/9)}",
      journal = {\apj},
         year = 2001,
        month = jun,
       volume = {554},
       number = {2},
        pages = {1035-1043},
          doi = {10.1086/321397},
archivePrefix = {arXiv},
       eprint = {astro-ph/0102256},
 primaryClass = {astro-ph},
       adsurl = {https://ui.adsabs.harvard.edu/abs/2001ApJ...554.1035F}
}

@ARTICLE{roberts2002,
       author = {{Roberts}, T.~P. and {Warwick}, R.~S. and {Ward}, M.~J. and {Murray}, S.~S.},
        title = "{A Chandra observation of the interacting pair of galaxies NGC 4485/4490}",
      journal = {\mnras},
         year = 2002,
        month = dec,
       volume = {337},
       number = {2},
        pages = {677-692},
          doi = {10.1046/j.1365-8711.2002.05950.x},
archivePrefix = {arXiv},
       eprint = {astro-ph/0208196},
 primaryClass = {astro-ph},
       adsurl = {https://ui.adsabs.harvard.edu/abs/2002MNRAS.337..677R}
}

@ARTICLE{gladstone2009,
       author = {{Gladstone}, Jeanette C. and {Roberts}, Timothy P.},
        title = "{The ultraluminous X-ray source population of NGC 4485/4490}",
      journal = {\mnras},
         year = 2009,
        month = jul,
       volume = {397},
       number = {1},
        pages = {124-134},
          doi = {10.1111/j.1365-2966.2009.14937.x},
archivePrefix = {arXiv},
       eprint = {0904.3852},
 primaryClass = {astro-ph.HE},
       adsurl = {https://ui.adsabs.harvard.edu/abs/2009MNRAS.397..124G}
}

@ARTICLE{swartz2008,
       author = {{Swartz}, Douglas A. and {Soria}, Roberto and {Tennant}, Allyn F.},
        title = "{Do Ultraluminous X-Ray Sources Exist in Dwarf Galaxies?}",
      journal = {\apj},
         year = 2008,
        month = sep,
       volume = {684},
       number = {1},
        pages = {282-286},
          doi = {10.1086/587776},
archivePrefix = {arXiv},
       eprint = {0803.1984},
 primaryClass = {astro-ph},
       adsurl = {https://ui.adsabs.harvard.edu/abs/2008ApJ...684..282S}
}

@ARTICLE{walton2011,
       author = {{Walton}, D.~J. and {Roberts}, T.~P. and {Mateos}, S. and {Heard}, V.},
        title = "{2XMM ultraluminous X-ray source candidates in nearby galaxies}",
      journal = {\mnras},
         year = 2011,
        month = sep,
       volume = {416},
       number = {3},
        pages = {1844-1861},
          doi = {10.1111/j.1365-2966.2011.19154.x},
archivePrefix = {arXiv},
       eprint = {1106.0197},
 primaryClass = {astro-ph.HE},
       adsurl = {https://ui.adsabs.harvard.edu/abs/2011MNRAS.416.1844W}
}

@ARTICLE{soria2006,
       author = {{Soria}, R. and {Fender}, R.~P. and {Hannikainen}, D.~C. and {Read}, A.~M. and {Stevens}, I.~R.},
        title = "{An ultraluminous X-ray microquasar in NGC5408?}",
      journal = {\mnras},
         year = 2006,
        month = jun,
       volume = {368},
       number = {4},
        pages = {1527-1539},
          doi = {10.1111/j.1365-2966.2006.10250.x},
archivePrefix = {arXiv},
       eprint = {astro-ph/0603029},
 primaryClass = {astro-ph},
       adsurl = {https://ui.adsabs.harvard.edu/abs/2006MNRAS.368.1527S}
}

@INPROCEEDINGS{soria2005,
       author = {{Soria}, R.},
        title = "{Runaway Core Collapse and Cluster Survival: Where are the Parent Clusters of ULXs?}",
    booktitle = {22nd Texas Symposium on Relativistic Astrophysics},
         year = 2005,
       editor = {{Chen}, Pisin and {Bloom}, Elliott and {Madejski}, Greg and {Patrosian}, Vahe},
        month = jan,
        pages = {517-522},
          doi = {10.48550/arXiv.astro-ph/0503340},
archivePrefix = {arXiv},
       eprint = {astro-ph/0503340},
 primaryClass = {astro-ph},
       adsurl = {https://ui.adsabs.harvard.edu/abs/2005tsra.conf..517S}
}

@ARTICLE{bregman2006,
       author = {{Bregman}, Joel N. and {Irwin}, Jimmy A. and {Seitzer}, Patrick and {Flores}, Matt},
        title = "{Galactic Globular Clusters with Luminous X-Ray Binaries}",
      journal = {\apj},
         year = 2006,
        month = mar,
       volume = {640},
       number = {1},
        pages = {282-287},
          doi = {10.1086/500037},
archivePrefix = {arXiv},
       eprint = {astro-ph/0511635},
 primaryClass = {astro-ph},
       adsurl = {https://ui.adsabs.harvard.edu/abs/2006ApJ...640..282B}
}

@ARTICLE{thygesen2023,
       author = {{Thygesen}, Erica and {Sun}, Yifan and {Huang}, Jeff and {Dage}, Kristen C. and {Zepf}, Stephen E. and {Kundu}, Arunav and {Haggard}, Daryl and {Maccarone}, Thomas J.},
        title = "{Globular cluster ultraluminous X-ray sources in the furthest early-type galaxies}",
      journal = {\mnras},
         year = 2023,
        month = jan,
       volume = {518},
       number = {3},
        pages = {3386-3396},
          doi = {10.1093/mnras/stac3244},
archivePrefix = {arXiv},
       eprint = {2211.07699},
 primaryClass = {astro-ph.HE},
       adsurl = {https://ui.adsabs.harvard.edu/abs/2023MNRAS.518.3386T}
}

@ARTICLE{brorby2014,
       author = {{Brorby}, M. and {Kaaret}, P. and {Prestwich}, A.},
        title = "{X-ray binary formation in low-metallicity blue compact dwarf galaxies}",
      journal = {\mnras},
         year = 2014,
        month = jul,
       volume = {441},
       number = {3},
        pages = {2346-2353},
          doi = {10.1093/mnras/stu736},
archivePrefix = {arXiv},
       eprint = {1404.3132},
 primaryClass = {astro-ph.GA},
       adsurl = {https://ui.adsabs.harvard.edu/abs/2014MNRAS.441.2346B}
}

@INPROCEEDINGS{elmellah2019,
       author = {{Mellah}, Ileyk El and {Sander}, Andreas A.~C. and {Sundqvist}, Jon O. and {Keppens}, Rony},
        title = "{Clumpy wind accretion in Supergiant X-ray Binaries}",
    booktitle = {High-mass X-ray Binaries: Illuminating the Passage from Massive Binaries to Merging Compact Objects},
         year = 2019,
       editor = {{Oskinova}, Lidia M. and {Bozzo}, Enrico and {Bulik}, Tomasz and {Gies}, Douglas R.},
       series = {IAU Symposium},
       volume = {346},
        month = dec,
        pages = {34-39},
          doi = {10.1017/S1743921318008414},
       adsurl = {https://ui.adsabs.harvard.edu/abs/2019IAUS..346...34M}
}

@ARTICLE{pinto2016,
       author = {{Pinto}, Ciro and {Middleton}, Matthew J. and {Fabian}, Andrew C.},
        title = "{Resolved atomic lines reveal outflows in two ultraluminous X-ray sources}",
      journal = {\nat},
         year = 2016,
        month = may,
       volume = {533},
       number = {7601},
        pages = {64-67},
          doi = {10.1038/nature17417},
archivePrefix = {arXiv},
       eprint = {1604.08593},
 primaryClass = {astro-ph.HE},
       adsurl = {https://ui.adsabs.harvard.edu/abs/2016Natur.533...64P}
}

@ARTICLE{kosec2018,
       author = {{Kosec}, P. and {Pinto}, C. and {Fabian}, A.~C. and {Walton}, D.~J.},
        title = "{Searching for outflows in ultraluminous X-ray sources through high-resolution X-ray spectroscopy}",
      journal = {\mnras},
         year = 2018,
        month = feb,
       volume = {473},
       number = {4},
        pages = {5680-5697},
          doi = {10.1093/mnras/stx2695},
archivePrefix = {arXiv},
       eprint = {1710.06438},
 primaryClass = {astro-ph.HE},
       adsurl = {https://ui.adsabs.harvard.edu/abs/2018MNRAS.473.5680K}
}

@ARTICLE{pinto2021,
       author = {{Pinto}, C. and {Soria}, R. and {Walton}, D.~J. and {D'A{\`\i}}, A. and {Pintore}, F. and {Kosec}, P. and {Alston}, W.~N. and {Fuerst}, F. and {Middleton}, M.~J. and {Roberts}, T.~P. and {Del Santo}, M. and {Barret}, D. and {Ambrosi}, E. and {Robba}, A. and {Earnshaw}, H. and {Fabian}, A.~C.},
        title = "{XMM-Newton campaign on the ultraluminous X-ray source NGC 247 ULX-1: outflows}",
      journal = {\mnras},
         year = 2021,
        month = aug,
       volume = {505},
       number = {4},
        pages = {5058-5074},
          doi = {10.1093/mnras/stab1648},
archivePrefix = {arXiv},
       eprint = {2104.11164},
 primaryClass = {astro-ph.HE},
       adsurl = {https://ui.adsabs.harvard.edu/abs/2021MNRAS.505.5058P}
}

@ARTICLE{barra2022,
       author = {{Barra}, F. and {Pinto}, C. and {Walton}, D.~J. and {Kosec}, P. and {D'A{\`\i}}, A. and {Di Salvo}, T. and {Del Santo}, M. and {Earnshaw}, H. and {Fabian}, A.~C. and {Fuerst}, F. and {Marino}, A. and {Pintore}, F. and {Robba}, A. and {Roberts}, T.~P.},
        title = "{Unveiling the disc structure in ultraluminous X-ray source NGC 55 ULX-1}",
      journal = {\mnras},
         year = 2022,
        month = nov,
       volume = {516},
       number = {3},
        pages = {3972-3983},
          doi = {10.1093/mnras/stac2453},
archivePrefix = {arXiv},
       eprint = {2207.12870},
 primaryClass = {astro-ph.HE},
       adsurl = {https://ui.adsabs.harvard.edu/abs/2022MNRAS.516.3972B}
}

@ARTICLE{robba2021,
       author = {{Robba}, A. and {Pinto}, C. and {Walton}, D.~J. and {Soria}, R. and {Kosec}, P. and {Pintore}, F. and {Roberts}, T.~P. and {Alston}, W.~N. and {Middleton}, M. and {Cusumano}, G. and {Earnshaw}, H.~P. and {F{\"u}rst}, F. and {Sathyaprakash}, R. and {Kyritsis}, E. and {Fabian}, A.~C.},
        title = "{Broadband X-ray spectral variability of the pulsing ULX NGC 1313 X-2}",
      journal = {\aap},
         year = 2021,
        month = aug,
       volume = {652},
          eid = {A118},
        pages = {A118},
          doi = {10.1051/0004-6361/202140884},
archivePrefix = {arXiv},
       eprint = {2106.04501},
 primaryClass = {astro-ph.HE},
       adsurl = {https://ui.adsabs.harvard.edu/abs/2021A&A...652A.118R}
}

@ARTICLE{bachetti2014,
       author = {{Bachetti}, M. and {Harrison}, F.~A. and {Walton}, D.~J. and {Grefenstette}, B.~W. and {Chakrabarty}, D. and {F{\"u}rst}, F. and {Barret}, D. and {Beloborodov}, A. and {Boggs}, S.~E. and {Christensen}, F.~E. and {Craig}, W.~W. and {Fabian}, A.~C. and {Hailey}, C.~J. and {Hornschemeier}, A. and {Kaspi}, V. and {Kulkarni}, S.~R. and {Maccarone}, T. and {Miller}, J.~M. and {Rana}, V. and {Stern}, D. and {Tendulkar}, S.~P. and {Tomsick}, J. and {Webb}, N.~A. and {Zhang}, W.~W.},
        title = "{An ultraluminous X-ray source powered by an accreting neutron star}",
      journal = {\nat},
         year = 2014,
        month = oct,
       volume = {514},
       number = {7521},
        pages = {202-204},
          doi = {10.1038/nature13791},
archivePrefix = {arXiv},
       eprint = {1410.3590},
 primaryClass = {astro-ph.HE},
       adsurl = {https://ui.adsabs.harvard.edu/abs/2014Natur.514..202B}
}

@ARTICLE{zampieri_2003,
       author = {{Zampieri}, L. and {Pastorello}, A. and {Turatto}, M. and {Cappellaro}, E. and {Benetti}, S. and {Altavilla}, G. and {Mazzali}, P. and {Hamuy}, M.},
        title = "{Core-collapse supernovae and evidence for black hole formation}",
      journal = {\memsai},
         year = 2003,
        month = jan,
       volume = {74},
        pages = {526},
       adsurl = {https://ui.adsabs.harvard.edu/abs/2003MmSAI..74..526Z}
}

@ARTICLE{israel2017a,
       author = {{Israel}, Gian Luca and {Belfiore}, Andrea and {Stella}, Luigi and {Esposito}, Paolo and {Casella}, Piergiorgio and {De Luca}, Andrea and {Marelli}, Martino and {Papitto}, Alessandro and {Perri}, Matteo and {Puccetti}, Simonetta and {Castillo}, Guillermo A. Rodr{\'\i}guez and {Salvetti}, David and {Tiengo}, Andrea and {Zampieri}, Luca and {D'Agostino}, Daniele and {Greiner}, Jochen and {Haberl}, Frank and {Novara}, Giovanni and {Salvaterra}, Ruben and {Turolla}, Roberto and {Watson}, Mike and {Wilms}, Joern and {Wolter}, Anna},
        title = "{An accreting pulsar with extreme properties drives an ultraluminous x-ray source in NGC 5907}",
      journal = {Science},
         year = 2017,
        month = feb,
       volume = {355},
       number = {6327},
        pages = {817-819},
          doi = {10.1126/science.aai8635},
archivePrefix = {arXiv},
       eprint = {1609.07375},
 primaryClass = {astro-ph.HE},
       adsurl = {https://ui.adsabs.harvard.edu/abs/2017Sci...355..817I}
}

@ARTICLE{fuerst16,
       author = {{F{\"u}rst}, F. and {Walton}, D.~J. and {Harrison}, F.~A. and {Stern}, D. and {Barret}, D. and {Brightman}, M. and {Fabian}, A.~C. and {Grefenstette}, B. and {Madsen}, K.~K. and {Middleton}, M.~J. and {Miller}, J.~M. and {Pottschmidt}, K. and {Ptak}, A. and {Rana}, V. and {Webb}, N.},
        title = "{Discovery of Coherent Pulsations from the Ultraluminous X-Ray Source NGC 7793 P13}",
      journal = {\apjl},
         year = 2016,
        month = nov,
       volume = {831},
       number = {2},
          eid = {L14},
        pages = {L14},
          doi = {10.3847/2041-8205/831/2/L14},
archivePrefix = {arXiv},
       eprint = {1609.07129},
 primaryClass = {astro-ph.HE},
       adsurl = {https://ui.adsabs.harvard.edu/abs/2016ApJ...831L..14F}
}

@ARTICLE{israel2017b,
       author = {{Israel}, G.~L. and {Papitto}, A. and {Esposito}, P. and {Stella}, L. and {Zampieri}, L. and {Belfiore}, A. and {Rodr{\'\i}guez Castillo}, G.~A. and {De Luca}, A. and {Tiengo}, A. and {Haberl}, F. and {Greiner}, J. and {Salvaterra}, R. and {Sandrelli}, S. and {Lisini}, G.},
        title = "{Discovery of a 0.42-s pulsar in the ultraluminous X-ray source NGC 7793 P13}",
      journal = {\mnras},
         year = 2017,
        month = mar,
       volume = {466},
       number = {1},
        pages = {L48-L52},
          doi = {10.1093/mnrasl/slw218},
archivePrefix = {arXiv},
       eprint = {1609.06538},
 primaryClass = {astro-ph.HE},
       adsurl = {https://ui.adsabs.harvard.edu/abs/2017MNRAS.466L..48I}
}

@ARTICLE{carpano2018,
       author = {{Carpano}, S. and {Haberl}, F. and {Maitra}, C. and {Vasilopoulos}, G.},
        title = "{Discovery of pulsations from NGC 300 ULX1 and its fast period evolution}",
      journal = {\mnras},
         year = 2018,
        month = may,
       volume = {476},
       number = {1},
        pages = {L45-L49},
          doi = {10.1093/mnrasl/sly030},
archivePrefix = {arXiv},
       eprint = {1802.10341},
 primaryClass = {astro-ph.HE},
       adsurl = {https://ui.adsabs.harvard.edu/abs/2018MNRAS.476L..45C}
}

@ARTICLE{sathyaprakash2019,
       author = {{Sathyaprakash}, R. and {Roberts}, T.~P. and {Walton}, D.~J. and {Fuerst}, F. and {Bachetti}, M. and {Pinto}, C. and {Alston}, W.~N. and {Earnshaw}, H.~P. and {Fabian}, A.~C. and {Middleton}, M.~J. and {Soria}, R.},
        title = "{The discovery of weak coherent pulsations in the ultraluminous X-ray source NGC 1313 X-2}",
      journal = {\mnras},
         year = 2019,
        month = sep,
       volume = {488},
       number = {1},
        pages = {L35-L40},
          doi = {10.1093/mnrasl/slz086},
archivePrefix = {arXiv},
       eprint = {1906.00640},
 primaryClass = {astro-ph.HE},
       adsurl = {https://ui.adsabs.harvard.edu/abs/2019MNRAS.488L..35S}
}

@ARTICLE{misra2023,
       author = {{Misra}, Devina and {Kovlakas}, Konstantinos and {Fragos}, Tassos and {Lazzarini}, Margaret and {Bavera}, Simone S. and {Lehmer}, Bret D. and {Zezas}, Andreas and {Zapartas}, Emmanouil and {Xing}, Zepei and {Andrews}, Jeff J. and {Dotter}, Aaron and {Rocha}, Kyle Akira and {Srivastava}, Philipp M. and {Sun}, Meng},
        title = "{X-ray luminosity function of high-mass X-ray binaries: Studying the signatures of different physical processes using detailed binary evolution calculations}",
      journal = {\aap},
         year = 2023,
        month = apr,
       volume = {672},
          eid = {A99},
        pages = {A99},
          doi = {10.1051/0004-6361/202244929},
archivePrefix = {arXiv},
       eprint = {2209.05505},
 primaryClass = {astro-ph.HE},
       adsurl = {https://ui.adsabs.harvard.edu/abs/2023A&A...672A..99M}
}

@ARTICLE{athena_code_dev,
       author = {{Qiao}, Erlin and {Wu}, Yongxin and {Lin}, Yiyang and {Guo}, Meng and {Liu}, Jifeng and {Guo}, Chenlei and {Jin}, Chichuan and {Jiang}, Ning},
        title = "{Early evolution of super-Eddington accretion flow in tidal disruption events}",
      journal = {\mnras},
         year = 2025,
        month = jun,
       volume = {539},
       number = {4},
        pages = {3473-3488},
          doi = {10.1093/mnras/staf719},
archivePrefix = {arXiv},
       eprint = {2505.02434},
 primaryClass = {astro-ph.HE},
       adsurl = {https://ui.adsabs.harvard.edu/abs/2025MNRAS.539.3473Q}
}

@ARTICLE{rodriguezcastillo2019,
       author = {{Rodr{\'\i}guez Castillo}, G.~A. and {Israel}, G.~L. and {Belfiore}, A. and {Bernardini}, F. and {Esposito}, P. and {Pintore}, F. and {De Luca}, A. and {Papitto}, A. and {Stella}, L. and {Tiengo}, A. and {Zampieri}, L. and {Bachetti}, M. and {Brightman}, M. and {Casella}, P. and {D'Agostino}, D. and {Dall'Osso}, S. and {Earnshaw}, H.~P. and {F{\"u}rst}, F. and {Haberl}, F. and {Harrison}, F.~A. and {Mapelli}, M. and {Marelli}, M. and {Middleton}, M. and {Pinto}, C. and {Roberts}, T.~P. and {Salvaterra}, R. and {Turolla}, R. and {Walton}, D.~J. and {Wolter}, A.},
        title = "{Discovery of a 2.8 s Pulsar in a 2 Day Orbit High-mass X-Ray Binary Powering the Ultraluminous X-Ray Source ULX-7 in M51}",
      journal = {\apj},
         year = 2020,
        month = may,
       volume = {895},
       number = {1},
          eid = {60},
        pages = {60},
          doi = {10.3847/1538-4357/ab8a44},
archivePrefix = {arXiv},
       eprint = {1906.04791},
 primaryClass = {astro-ph.HE},
       adsurl = {https://ui.adsabs.harvard.edu/abs/2020ApJ...895...60R}
}

@ARTICLE{vasilopoulos2020,
       author = {{Vasilopoulos}, G. and {Ray}, P.~S. and {Gendreau}, K.~C. and {Jenke}, P.~A. and {Jaisawal}, G.~K. and {Wilson-Hodge}, C.~A. and {Strohmayer}, T.~E. and {Altamirano}, D. and {Iwakiri}, W.~B. and {Wolff}, M.~T. and {Guillot}, S. and {Malacaria}, C. and {Stevens}, A.~L.},
        title = "{The 2019 super-Eddington outburst of RX J0209.6-7427: detection of pulsations and constraints on the magnetic field strength}",
      journal = {\mnras},
         year = 2020,
        month = jun,
       volume = {494},
       number = {4},
        pages = {5350-5359},
          doi = {10.1093/mnras/staa991},
archivePrefix = {arXiv},
       eprint = {2004.03022},
 primaryClass = {astro-ph.HE},
       adsurl = {https://ui.adsabs.harvard.edu/abs/2020MNRAS.494.5350V}
}

@ARTICLE{kennea2017,
       author = {{Kennea}, J.~A. and {Lien}, A.~Y. and {Krimm}, H.~A. and {Cenko}, S.~B. and {Siegel}, M.~H.},
        title = "{Swift J0243.6+6124: Swift discovery of an accreting NS transient}",
      journal = {The Astronomer's Telegram},
         year = 2017,
        month = oct,
       volume = {10809},
        pages = {1},
       adsurl = {https://ui.adsabs.harvard.edu/abs/2017ATel10809....1K}
}

@ARTICLE{brightman2019,
       author = {{Brightman}, Murray and {Harrison}, Fiona A. and {Bachetti}, Matteo and {Xu}, Yanjun and {F{\"u}rst}, Felix and {Walton}, Dominic J. and {Ptak}, Andrew and {Yukita}, Mihoko and {Zezas}, Andreas},
        title = "{A {\ensuremath{\sim}}60 day Super-orbital Period Originating from the Ultraluminous X-Ray Pulsar in M82}",
      journal = {\apj},
         year = 2019,
        month = mar,
       volume = {873},
       number = {2},
          eid = {115},
        pages = {115},
          doi = {10.3847/1538-4357/ab0215},
archivePrefix = {arXiv},
       eprint = {1901.10491},
 primaryClass = {astro-ph.HE},
       adsurl = {https://ui.adsabs.harvard.edu/abs/2019ApJ...873..115B}
}

@ARTICLE{urquhart2016,
       author = {{Urquhart}, R. and {Soria}, R.},
        title = "{Two Eclipsing Ultraluminous X-Ray Sources in M51}",
      journal = {\apj},
         year = 2016,
        month = nov,
       volume = {831},
       number = {1},
          eid = {56},
        pages = {56},
          doi = {10.3847/0004-637X/831/1/56},
archivePrefix = {arXiv},
       eprint = {1608.01111},
 primaryClass = {astro-ph.HE},
       adsurl = {https://ui.adsabs.harvard.edu/abs/2016ApJ...831...56U}
}

@ARTICLE{Shigeyuki2022,
       author = {{Karino}, Shigeyuki},
        title = "{Characteristics and evolution of Be-type high-mass X-ray binaries as potential ultraluminous X-ray sources}",
      journal = {\mnras},
         year = 2022,
        month = jul,
       volume = {514},
       number = {1},
        pages = {191-199},
          doi = {10.1093/mnras/stac1334},
archivePrefix = {arXiv},
       eprint = {2205.04607},
 primaryClass = {astro-ph.SR},
       adsurl = {https://ui.adsabs.harvard.edu/abs/2022MNRAS.514..191K}
}

@ARTICLE{earnshaw2018,
       author = {{Earnshaw}, H.~P. and {Roberts}, T.~P. and {Sathyaprakash}, R.},
        title = "{Searching for propeller-phase ULXs in the XMM-Newton Serendipitous Source Catalogue}",
      journal = {\mnras},
         year = 2018,
        month = may,
       volume = {476},
       number = {3},
        pages = {4272-4277},
          doi = {10.1093/mnras/sty501},
archivePrefix = {arXiv},
       eprint = {1802.07753},
 primaryClass = {astro-ph.HE},
       adsurl = {https://ui.adsabs.harvard.edu/abs/2018MNRAS.476.4272E}
}

@ARTICLE{earnshaw2024,
       author = {{Earnshaw}, Hannah P. and {Patti}, Gauri and {Brightman}, Murray and {Sathyaprakash}, Rajath and {Walton}, Dominic J. and {Fuerst}, Felix and {Roberts}, Timothy P. and {Harrison}, Fiona A.},
        title = "{The long-term variability of a population of ULXs monitored by Chandra}",
      journal = {arXiv e-prints},
         year = 2024,
        month = nov,
          eid = {arXiv:2411.07459},
        pages = {arXiv:2411.07459},
archivePrefix = {arXiv},
       eprint = {2411.07459},
 primaryClass = {astro-ph.HE},
       adsurl = {https://ui.adsabs.harvard.edu/abs/2024arXiv241107459E}
}

@ARTICLE{tsygankov2016,
       author = {{Tsygankov}, Sergey S. and {Mushtukov}, Alexander A. and {Suleimanov}, Valery F. and {Poutanen}, Juri},
        title = "{Propeller effect in action in the ultraluminous accreting magnetar M82 X-2}",
      journal = {\mnras},
         year = 2016,
        month = mar,
       volume = {457},
       number = {1},
        pages = {1101-1106},
          doi = {10.1093/mnras/stw046},
archivePrefix = {arXiv},
       eprint = {1507.08288},
 primaryClass = {astro-ph.HE},
       adsurl = {https://ui.adsabs.harvard.edu/abs/2016MNRAS.457.1101T}
}

@INPROCEEDINGS{grise09,
       author = {{Gris{\'e}}, Fabien and {Pakull}, Manfred W. and {Soria}, Roberto and {Motch}, Christian},
        title = "{The ULX NGC 1313 X-2: an Optical Study Revealing an Interesting Behavior}",
    booktitle = {SIMBOL-X: Focusing on the Hard X-ray Universe},
         year = 2009,
       editor = {{Rodriguez}, J{\'e}r{\^o}me and {Ferrando}, Phillippe},
       series = {American Institute of Physics Conference Series},
       volume = {1126},
        month = may,
    publisher = {AIP},
        pages = {201-203},
          doi = {10.1063/1.3149412},
archivePrefix = {arXiv},
       eprint = {0902.4431},
 primaryClass = {astro-ph.HE},
       adsurl = {https://ui.adsabs.harvard.edu/abs/2009AIPC.1126..201G}
}

@ARTICLE{grise11,
       author = {{Gris{\'e}}, F. and {Kaaret}, P. and {Pakull}, M.~W. and {Motch}, C.},
        title = "{Optical Properties of the Ultraluminous X-Ray Source Holmberg IX X-1 and Its Stellar Environment}",
      journal = {\apj},
         year = 2011,
        month = jun,
       volume = {734},
       number = {1},
          eid = {23},
        pages = {23},
          doi = {10.1088/0004-637X/734/1/23},
archivePrefix = {arXiv},
       eprint = {1104.5523},
 primaryClass = {astro-ph.HE},
       adsurl = {https://ui.adsabs.harvard.edu/abs/2011ApJ...734...23G}
}

@ARTICLE{grise12,
       author = {{Gris{\'e}}, F. and {Kaaret}, P. and {Corbel}, S. and {Feng}, H. and {Cseh}, D. and {Tao}, L.},
        title = "{Optical Emission of the Ultraluminous X-Ray Source NGC 5408 X-1: Donor Star or Irradiated Accretion Disk?}",
      journal = {\apj},
         year = 2012,
        month = feb,
       volume = {745},
       number = {2},
          eid = {123},
        pages = {123},
          doi = {10.1088/0004-637X/745/2/123},
archivePrefix = {arXiv},
       eprint = {1109.4423},
 primaryClass = {astro-ph.HE},
       adsurl = {https://ui.adsabs.harvard.edu/abs/2012ApJ...745..123G}
}

@INPROCEEDINGS{pakull2006,
       author = {{Pakull}, Manfred W. and {Gris{\'e}}, Fabien and {Motch}, Christian},
        title = "{Ultraluminous X-ray Sources: Bubbles and Optical Counterparts}",
    booktitle = {Populations of High Energy Sources in Galaxies},
         year = 2006,
       editor = {{Meurs}, E.~J.~A. and {Fabbiano}, G.},
       series = {IAU Symposium},
       volume = {230},
        month = jan,
        pages = {293-297},
          doi = {10.1017/S1743921306008489},
archivePrefix = {arXiv},
       eprint = {astro-ph/0603771},
 primaryClass = {astro-ph},
       adsurl = {https://ui.adsabs.harvard.edu/abs/2006IAUS..230..293P}
}

@INPROCEEDINGS{pakull2008,
       author = {{Pakull}, Manfred W. and {Gris{\'e}}, Fabien},
        title = "{Ultraluminous X-ray Sources: Beambags and Optical Counterparts}",
    booktitle = {A Population Explosion: The Nature \& Evolution of X-ray Binaries in Diverse Environments},
         year = 2008,
       editor = {{Bandyopadhyay}, Reba M. and {Wachter}, Stefanie and {Gelino}, Dawn and {Gelino}, Christopher R.},
       series = {American Institute of Physics Conference Series},
       volume = {1010},
        month = may,
    publisher = {AIP},
        pages = {303-307},
          doi = {10.1063/1.2945062},
archivePrefix = {arXiv},
       eprint = {0803.4345},
 primaryClass = {astro-ph},
       adsurl = {https://ui.adsabs.harvard.edu/abs/2008AIPC.1010..303P}
}

@ARTICLE{abolmasov2008,
       author = {{Abolmasov}, Pavel and {Fabrika}, S. and {Sholukhova}, O. and {Kotani}, Taro},
        title = "{Optical Spectroscopy of the ULX-Associated Nebula MF16}",
      journal = {arXiv e-prints},
         year = 2008,
        month = sep,
          eid = {arXiv:0809.0409},
        pages = {arXiv:0809.0409},
          doi = {10.48550/arXiv.0809.0409},
archivePrefix = {arXiv},
       eprint = {0809.0409},
 primaryClass = {astro-ph},
       adsurl = {https://ui.adsabs.harvard.edu/abs/2008arXiv0809.0409A}
}

@ARTICLE{kaaret2010,
       author = {{Kaaret}, Philip and {Feng}, Hua and {Wong}, Diane S. and {Tao}, Lian},
        title = "{Direct Detection of an Ultraluminous Ultraviolet Source}",
      journal = {\apjl},
         year = 2010,
        month = may,
       volume = {714},
       number = {1},
        pages = {L167-L170},
          doi = {10.1088/2041-8205/714/1/L167},
archivePrefix = {arXiv},
       eprint = {1003.4257},
 primaryClass = {astro-ph.HE},
       adsurl = {https://ui.adsabs.harvard.edu/abs/2010ApJ...714L.167K}
}

@ARTICLE{gurpide2024a,
       author = {{G{\'u}rpide}, A. and {Castro Segura}, N. and {Soria}, R. and {Middleton}, M.},
        title = "{Absence of nebular He II {\ensuremath{\lambda}}4686 constrains the UV emission from the ultraluminous X-ray pulsar NGC 1313 X-2}",
      journal = {\mnras},
         year = 2024,
        month = jul,
       volume = {531},
       number = {3},
        pages = {3118-3135},
          doi = {10.1093/mnras/stae1336},
archivePrefix = {arXiv},
       eprint = {2405.13714},
 primaryClass = {astro-ph.HE},
       adsurl = {https://ui.adsabs.harvard.edu/abs/2024MNRAS.531.3118G}
}

@ARTICLE{gurpide2024b,
       author = {{G{\'u}rpide}, A. and {Castro Segura}, N.},
        title = "{Quasi-isotropic UV emission in the ULX NGC 1313 X-1}",
      journal = {\mnras},
         year = 2024,
        month = aug,
       volume = {532},
       number = {2},
        pages = {1459-1485},
          doi = {10.1093/mnras/stae1329},
archivePrefix = {arXiv},
       eprint = {2405.14512},
 primaryClass = {astro-ph.HE},
       adsurl = {https://ui.adsabs.harvard.edu/abs/2024MNRAS.532.1459G}
}

@ARTICLE{beuchert2024,
       author = {{Beuchert}, Tobias and {Middleton}, Matthew J. and {Soria}, Roberto and {Miller-Jones}, James C.~A. and {Dauser}, Thomas and {Roberts}, Timothy P. and {Sathyaprakash}, Rajath and {Markoff}, Sera},
        title = "{Exploring the case for hard-X-ray beaming in NGC 6946 X-1}",
      journal = {\mnras},
         year = 2024,
        month = oct,
       volume = {534},
       number = {1},
        pages = {645-654},
          doi = {10.1093/mnras/stae1975},
archivePrefix = {arXiv},
       eprint = {2408.07751},
 primaryClass = {astro-ph.HE},
       adsurl = {https://ui.adsabs.harvard.edu/abs/2024MNRAS.534..645B}
}

@ARTICLE{cseh2014,
       author = {{Cseh}, D. and {Kaaret}, P. and {Corbel}, S. and {Grise}, F. and {Lang}, C. and {Kording}, E. and {Falcke}, H. and {Jonker}, P.~G. and {Miller-Jones}, J.~C.~A. and {Farrell}, S. and {Yang}, Y.~J. and {Paragi}, Z. and {Frey}, S.},
        title = "{Unveiling recurrent jets of the ULX Holmberg II X-1: evidence for a massive stellar-mass black hole?}",
      journal = {\mnras},
         year = 2014,
        month = mar,
       volume = {439},
        pages = {L1-L5},
          doi = {10.1093/mnrasl/slt166},
archivePrefix = {arXiv},
       eprint = {1311.4867},
 primaryClass = {astro-ph.HE},
       adsurl = {https://ui.adsabs.harvard.edu/abs/2014MNRAS.439L...1C}
}

@ARTICLE{pakull2010,
       author = {{Pakull}, Manfred W. and {Soria}, Roberto and {Motch}, Christian},
        title = "{A 300-parsec-long jet-inflated bubble around a powerful microquasar in the galaxy NGC 7793}",
      journal = {\nat},
         year = 2010,
        month = jul,
       volume = {466},
       number = {7303},
        pages = {209-212},
          doi = {10.1038/nature09168},
       adsurl = {https://ui.adsabs.harvard.edu/abs/2010Natur.466..209P}
}

@ARTICLE{mezcua2015,
       author = {{Mezcua}, M. and {Roberts}, T.~P. and {Lobanov}, A.~P. and {Sutton}, A.~D.},
        title = "{The powerful jet of an off-nuclear intermediate-mass black hole in the spiral galaxy NGC 2276}",
      journal = {\mnras},
         year = 2015,
        month = apr,
       volume = {448},
       number = {2},
        pages = {1893-1899},
          doi = {10.1093/mnras/stv143},
archivePrefix = {arXiv},
       eprint = {1501.04897},
 primaryClass = {astro-ph.GA},
       adsurl = {https://ui.adsabs.harvard.edu/abs/2015MNRAS.448.1893M}
}

@ARTICLE{Koljonen2021,
       author = {{Koljonen}, K.~I.~I. and {Hovatta}, T.},
        title = "{ALMA/NICER observations of GRS 1915+105 indicate a return to a hard state}",
      journal = {\aap},
         year = 2021,
        month = mar,
       volume = {647},
          eid = {A173},
        pages = {A173},
          doi = {10.1051/0004-6361/202039581},
archivePrefix = {arXiv},
       eprint = {2102.00693},
 primaryClass = {astro-ph.HE},
       adsurl = {https://ui.adsabs.harvard.edu/abs/2021A&A...647A.173K}
}

@ARTICLE{motch2014,
       author = {{Motch}, C. and {Pakull}, M.~W. and {Soria}, R. and {Gris{\'e}}, F. and {Pietrzy{\'n}ski}, G.},
        title = "{A mass of less than 15 solar masses for the black hole in an ultraluminous X-ray source}",
      journal = {\nat},
         year = 2014,
        month = oct,
       volume = {514},
       number = {7521},
        pages = {198-201},
          doi = {10.1038/nature13730},
archivePrefix = {arXiv},
       eprint = {1410.4250},
 primaryClass = {astro-ph.HE},
       adsurl = {https://ui.adsabs.harvard.edu/abs/2014Natur.514..198M}
}

@ARTICLE{Liu2013,
       author = {{Liu}, Ji-Feng and {Bregman}, Joel N. and {Bai}, Yu and {Justham}, Stephen and {Crowther}, Paul},
        title = "{Puzzling accretion onto a black hole in the ultraluminous X-ray source M 101 ULX-1}",
      journal = {\nat},
         year = 2013,
        month = nov,
       volume = {503},
       number = {7477},
        pages = {500-503},
          doi = {10.1038/nature12762},
archivePrefix = {arXiv},
       eprint = {1312.0337},
 primaryClass = {astro-ph.HE},
       adsurl = {https://ui.adsabs.harvard.edu/abs/2013Natur.503..500L}
}

@ARTICLE{heida2019,
       author = {{Heida}, M. and {Lau}, R.~M. and {Davies}, B. and {Brightman}, M. and {F{\"u}rst}, F. and {Grefenstette}, B.~W. and {Kennea}, J.~A. and {Tramper}, F. and {Walton}, D.~J. and {Harrison}, F.~A.},
        title = "{Discovery of a Red Supergiant Donor Star in SN2010da/NGC 300 ULX-1}",
      journal = {\apjl},
         year = 2019,
        month = oct,
       volume = {883},
       number = {2},
          eid = {L34},
        pages = {L34},
          doi = {10.3847/2041-8213/ab4139},
archivePrefix = {arXiv},
       eprint = {1909.02171},
 primaryClass = {astro-ph.HE},
       adsurl = {https://ui.adsabs.harvard.edu/abs/2019ApJ...883L..34H}
}

@ARTICLE{heida2019a,
       author = {{Heida}, M. and {Harrison}, F.~A. and {Brightman}, M. and {F{\"u}rst}, F. and {Stern}, D. and {Walton}, D.~J.},
        title = "{Searching for the Donor Stars of ULX Pulsars}",
      journal = {\apj},
         year = 2019,
        month = feb,
       volume = {871},
       number = {2},
          eid = {231},
        pages = {231},
          doi = {10.3847/1538-4357/aafa77},
archivePrefix = {arXiv},
       eprint = {1901.03776},
 primaryClass = {astro-ph.HE},
       adsurl = {https://ui.adsabs.harvard.edu/abs/2019ApJ...871..231H}
}

@ARTICLE{roberts2008,
       author = {{Roberts}, T.~P. and {Levan}, A.~J. and {Goad}, M.~R.},
        title = "{New Hubble Space Telescope imaging of the counterparts to six ultraluminous X-ray sources}",
      journal = {\mnras},
         year = 2008,
        month = jun,
       volume = {387},
       number = {1},
        pages = {73-78},
          doi = {10.1111/j.1365-2966.2008.13293.x},
archivePrefix = {arXiv},
       eprint = {0803.4470},
 primaryClass = {astro-ph},
       adsurl = {https://ui.adsabs.harvard.edu/abs/2008MNRAS.387...73R}
}

@ARTICLE{gladstone2013,
       author = {{Gladstone}, Jeanette C. and {Copperwheat}, Chris and {Heinke}, Craig O. and {Roberts}, Timothy P. and {Cartwright}, Taylor F. and {Levan}, Andrew J. and {Goad}, Mike R.},
        title = "{Optical Counterparts of the Nearest Ultraluminous X-Ray Sources}",
      journal = {\apjs},
         year = 2013,
        month = jun,
       volume = {206},
       number = {2},
          eid = {14},
        pages = {14},
          doi = {10.1088/0067-0049/206/2/14},
archivePrefix = {arXiv},
       eprint = {1303.1213},
 primaryClass = {astro-ph.HE},
       adsurl = {https://ui.adsabs.harvard.edu/abs/2013ApJS..206...14G}
}

@ARTICLE{allak2024,
       author = {{Allak}, Sinan},
        title = "{The first glimpse of ULXs through the near-infrared images captured by the JWST}",
      journal = {\mnras},
         year = 2024,
        month = jan,
       volume = {527},
       number = {2},
        pages = {2599-2611},
          doi = {10.1093/mnras/stad3332},
archivePrefix = {arXiv},
       eprint = {2306.11163},
 primaryClass = {astro-ph.HE},
       adsurl = {https://ui.adsabs.harvard.edu/abs/2024MNRAS.527.2599A}
}

@ARTICLE{Tao2011,
       author = {{Tao}, Lian and {Feng}, Hua and {Gris{\'e}}, Fabien and {Kaaret}, Philip},
        title = "{Compact Optical Counterparts of Ultraluminous X-Ray Sources}",
      journal = {\apj},
         year = 2011,
        month = aug,
       volume = {737},
       number = {2},
          eid = {81},
        pages = {81},
          doi = {10.1088/0004-637X/737/2/81},
archivePrefix = {arXiv},
       eprint = {1106.0315},
 primaryClass = {astro-ph.HE},
       adsurl = {https://ui.adsabs.harvard.edu/abs/2011ApJ...737...81T}
}

@ARTICLE{heida2014,
       author = {{Heida}, M. and {Jonker}, P.~G. and {Torres}, M.~A.~P. and {Kool}, E. and {Servillat}, M. and {Roberts}, T.~P. and {Groot}, P.~J. and {Walton}, D.~J. and {Moon}, D. -S. and {Harrison}, F.~A.},
        title = "{Near-infrared counterparts of ultraluminous X-ray sources}",
      journal = {\mnras},
         year = 2014,
        month = aug,
       volume = {442},
       number = {2},
        pages = {1054-1067},
          doi = {10.1093/mnras/stu928},
archivePrefix = {arXiv},
       eprint = {1405.1733},
 primaryClass = {astro-ph.HE},
       adsurl = {https://ui.adsabs.harvard.edu/abs/2014MNRAS.442.1054H}
}

@ARTICLE{heida2016,
       author = {{Heida}, M. and {Jonker}, P.~G. and {Torres}, M.~A.~P. and {Roberts}, T.~P. and {Walton}, D.~J. and {Moon}, D. -S. and {Stern}, D. and {Harrison}, F.~A.},
        title = "{Keck/MOSFIRE spectroscopy of five ULX counterparts}",
      journal = {\mnras},
         year = 2016,
        month = jun,
       volume = {459},
       number = {1},
        pages = {771-778},
          doi = {10.1093/mnras/stw695},
archivePrefix = {arXiv},
       eprint = {1603.07024},
 primaryClass = {astro-ph.HE},
       adsurl = {https://ui.adsabs.harvard.edu/abs/2016MNRAS.459..771H}
}

@ARTICLE{soria2017,
       author = {{Soria}, Roberto and {Musaeva}, Aina and {Wu}, Kinwah and {Zampieri}, Luca and {Federle}, Sara and {Urquhart}, Ryan and {van der Helm}, Edwin and {Farrell}, Sean},
        title = "{Outbursts of the intermediate-mass black hole HLX-1: a wind-instability scenario}",
      journal = {\mnras},
         year = 2017,
        month = jul,
       volume = {469},
       number = {1},
        pages = {886-905},
          doi = {10.1093/mnras/stx888},
archivePrefix = {arXiv},
       eprint = {1704.05468},
 primaryClass = {astro-ph.HE},
       adsurl = {https://ui.adsabs.harvard.edu/abs/2017MNRAS.469..886S}
}

@ARTICLE{webb2017,
       author = {{Webb}, N.~A. and {Gu{\'e}rou}, A. and {Ciambur}, B. and {Detoeuf}, A. and {Coriat}, M. and {Godet}, O. and {Barret}, D. and {Combes}, F. and {Contini}, T. and {Graham}, Alister W. and {Maccarone}, T.~J. and {Mrkalj}, M. and {Servillat}, M. and {Schroetter}, I. and {Wiersema}, K.},
        title = "{Understanding the environment around the intermediate mass black hole candidate ESO 243-49 HLX-1}",
      journal = {\aap},
         year = 2017,
        month = jun,
       volume = {602},
          eid = {A103},
        pages = {A103},
          doi = {10.1051/0004-6361/201630042},
archivePrefix = {arXiv},
       eprint = {1704.04434},
 primaryClass = {astro-ph.HE},
       adsurl = {https://ui.adsabs.harvard.edu/abs/2017A\&A...602A.103W}
}

@ARTICLE{lopez2017,
       author = {{L{\'o}pez}, K.~M. and {Heida}, M. and {Jonker}, P.~G. and {Torres}, M.~A.~P. and {Roberts}, T.~P. and {Walton}, D.~J. and {Moon}, D. -S. and {Harrison}, F.~A.},
        title = "{A systematic search for near-infrared counterparts of nearby ultraluminous X-ray sources (II)}",
      journal = {\mnras},
         year = 2017,
        month = jul,
       volume = {469},
       number = {1},
        pages = {671-682},
          doi = {10.1093/mnras/stx857},
archivePrefix = {arXiv},
       eprint = {1704.01068},
 primaryClass = {astro-ph.HE},
       adsurl = {https://ui.adsabs.harvard.edu/abs/2017MNRAS.469..671L}
}

@ARTICLE{lopez2020,
       author = {{L{\'o}pez}, K.~M. and {Heida}, M. and {Jonker}, P.~G. and {Torres}, M.~A.~P. and {Roberts}, T.~P. and {Walton}, D.~J. and {Moon}, D. -S. and {Harrison}, F.~A.},
        title = "{NIR counterparts to ULXs (III): completing the photometric survey and selected spectroscopic results}",
      journal = {\mnras},
         year = 2020,
        month = sep,
       volume = {497},
       number = {1},
        pages = {917-932},
          doi = {10.1093/mnras/staa1920},
archivePrefix = {arXiv},
       eprint = {2006.02795},
 primaryClass = {astro-ph.HE},
       adsurl = {https://ui.adsabs.harvard.edu/abs/2020MNRAS.497..917L}
}

@ARTICLE{lau2019,
       author = {{Lau}, Ryan M. and {Heida}, Marianne and {Walton}, Dominic J. and {Kasliwal}, Mansi M. and {Adams}, Scott M. and {Cody}, Ann Marie and {De}, Kishalay and {Gehrz}, Robert D. and {F{\"u}rst}, Felix and {Jencson}, Jacob E. and {Kennea}, Jamie A. and {Masci}, Frank},
        title = "{Uncovering Red and Dusty Ultraluminous X-Ray Sources with Spitzer}",
      journal = {\apj},
         year = 2019,
        month = jun,
       volume = {878},
       number = {1},
          eid = {71},
        pages = {71},
          doi = {10.3847/1538-4357/ab1b1c},
archivePrefix = {arXiv},
       eprint = {1904.09852},
 primaryClass = {astro-ph.HE},
       adsurl = {https://ui.adsabs.harvard.edu/abs/2019ApJ...878...71L}
}

@ARTICLE{dudik2016,
       author = {{Dudik}, R.~P. and {Berghea}, C.~T. and {Roberts}, T.~P. and {Gris{\'e}}, F. and {Singh}, A. and {Pagano}, R. and {Winter}, L.~M.},
        title = "{Spitzer IRAC Observations of IR Excess in Holmberg IX X-1: A Circumbinary Disk or a Variable Jet?}",
      journal = {\apj},
         year = 2016,
        month = nov,
       volume = {831},
       number = {1},
          eid = {88},
        pages = {88},
          doi = {10.3847/0004-637X/831/1/88},
archivePrefix = {arXiv},
       eprint = {1608.05348},
 primaryClass = {astro-ph.HE},
       adsurl = {https://ui.adsabs.harvard.edu/abs/2016ApJ...831...88D}
}

@ARTICLE{sathyaprakash2022,
       author = {{Sathyaprakash}, R. and {Roberts}, T.~P. and {Gris{\'e}}, F. and {Kaaret}, P. and {Ambrosi}, E. and {Done}, C. and {Gladstone}, J.~C. and {Kajava}, J.~J.~E. and {Soria}, R. and {Zampieri}, L.},
        title = "{A multi-wavelength view of distinct accretion regimes in the pulsating ultraluminous X-ray source NGC 1313 X-2}",
      journal = {\mnras},
         year = 2022,
        month = apr,
       volume = {511},
       number = {4},
        pages = {5346-5362},
          doi = {10.1093/mnras/stac402},
archivePrefix = {arXiv},
       eprint = {2202.06986},
 primaryClass = {astro-ph.HE},
       adsurl = {https://ui.adsabs.harvard.edu/abs/2022MNRAS.511.5346S}
}

@ARTICLE{pakull2002,
       author = {{Pakull}, Manfred W. and {Mirioni}, Laurent},
        title = "{Optical Counterparts of Ultraluminous X-Ray Sources}",
      journal = {arXiv e-prints},
         year = 2002,
        month = feb,
          eid = {astro-ph/0202488},
        pages = {astro-ph/0202488},
          doi = {10.48550/arXiv.astro-ph/0202488},
archivePrefix = {arXiv},
       eprint = {astro-ph/0202488},
 primaryClass = {astro-ph},
       adsurl = {https://ui.adsabs.harvard.edu/abs/2002astro.ph..2488P}
}

@ARTICLE{horne2021,
       author = {{Horne}, Keith and {De Rosa}, G. and {Peterson}, B.~M. and {Barth}, A.~J. and {Ely}, J. and {Fausnaugh}, M.~M. and {Kriss}, G.~A. and {Pei}, L. and {Bentz}, M.~C. and {Cackett}, E.~M. and {Edelson}, R. and {Eracleous}, M. and {Goad}, M.~R. and {Grier}, C.~J. and {Kaastra}, J. and {Kochanek}, C.~S. and {Krongold}, Y. and {Mathur}, S. and {Netzer}, H. and {Proga}, D. and {Tejos}, N. and {Vestergaard}, M. and {Villforth}, C. and {Adams}, S.~M. and {Anderson}, M.~D. and {Ar{\'e}valo}, P. and {Beatty}, T.~G. and {Bennert}, V.~N. and {Bigley}, A. and {Bisogni}, S. and {Borman}, G.~A. and {Boroson}, T.~A. and {Bottorff}, M.~C. and {Brandt}, W.~N. and {Breeveld}, A.~A. and {Brotherton}, M. and {Brown}, J.~E. and {Brown}, J.~S. and {Canalizo}, G. and {Carini}, M.~T. and {Clubb}, K.~I. and {Comerford}, J.~M. and {Corsini}, E.~M. and {Crenshaw}, D.~M. and {Croft}, S. and {Croxall}, K.~V. and {Dalla Bont{\`a}}, E. and {Deason}, A.~J. and {Dehghanian}, M. and {De Lorenzo-C{\'a}ceres}, A. and {Denney}, K.~D. and {Dietrich}, M. and {Done}, C. and {Efimova}, N.~V. and {Evans}, P.~A. and {Ferland}, G.~J. and {Filippenko}, A.~V. and {Flatland}, K. and {Fox}, O.~D. and {Gardner}, E. and {Gates}, E.~L. and {Gehrels}, N. and {Geier}, S. and {Gelbord}, J.~M. and {Gonzalez}, L. and {Gorjian}, V. and {Greene}, J.~E. and {Grupe}, D. and {Gupta}, A. and {Hall}, P.~B. and {Henderson}, C.~B. and {Hicks}, S. and {Holmbeck}, E. and {Holoien}, T.~W.-S. and {Hutchison}, T. and {Im}, M. and {Jensen}, J.~J. and {Johnson}, C.~A. and {Joner}, M.~D. and {Jones}, J. and {Kaspi}, S. and {Kelly}, P.~L. and {Kennea}, J.~A. and {Kim}, M. and {Kim}, S. and {Kim}, S.~C. and {King}, A. and {Klimanov}, S.~A. and {Korista}, K.~T. and {Lau}, M.~W. and {Lee}, J.~C. and {Leonard}, D.~C. and {Li}, Miao and {Lira}, P. and {Lochhaas}, C. and {Ma}, Zhiyuan and {MacInnis}, F. and {Malkan}, M.~A. and {Manne-Nicholas}, E.~R. and {Mauerhan}, J.~C. and {McGurk}, R. and {McHardy}, I.~M. and {Montuori}, C. and {Morelli}, L. and {Mosquera}, A. and {Mudd}, D. and {M{\"u}ller-S{\'a}nchez}, F. and {Nazarov}, S.~V. and {Norris}, R.~P. and {Nousek}, J.~A. and {Nguyen}, M.~L. and {Ochner}, P. and {Okhmat}, D.~N. and {Pancoast}, A. and {Papadakis}, I. and {Parks}, J.~R. and {Penny}, M.~T. and {Pizzella}, A. and {Pogge}, R.~W. and {Poleski}, R. and {Pott}, J.-U. and {Rafter}, S.~E. and {Rix}, H.-W. and {Runnoe}, J. and {Saylor}, D.~A. and {Schimoia}, J.~S. and {Schn{\"u}lle}, K. and {Scott}, B. and {Sergeev}, S.~G. and {Shappee}, B.~J. and {Shivvers}, I. and {Siegel}, M. and {Simonian}, G.~V. and {Siviero}, A. and {Skielboe}, A. and {Somers}, G. and {Spencer}, M. and {Starkey}, D. and {Stevens}, D.~J. and {Sung}, H.-I. and {Tayar}, J. and {Treu}, T. and {Turner}, C.~S. and {Uttley}, P. and {Van Saders}, J. and {Vican}, L. and {Villanueva}, Jr., S. and {Weiss}, Y. and {Woo}, J.-H. and {Yan}, H. and {Young}, S. and {Yuk}, H. and {Zheng}, W. and {Zhu}, W. and {Zu}, Y.},
        title = "{Space Telescope and Optical Reverberation Mapping Project. IX. Velocity-Delay Maps for Broad Emission Lines in NGC 5548}",
      journal = {\apj},
         year = 2021,
        month = feb,
       volume = {907},
       number = {2},
          eid = {76},
        pages = {76},
          doi = {10.3847/1538-4357/abce60},
archivePrefix = {arXiv},
       eprint = {2003.01448},
 primaryClass = {astro-ph.GA},
       adsurl = {https://ui.adsabs.harvard.edu/abs/2021ApJ...907...76H}
}

@ARTICLE{roberts2011,
       author = {{Roberts}, T.~P. and {Gladstone}, J.~C. and {Goulding}, A.~D. and {Swinbank}, A.~M. and {Ward}, M.~J. and {Goad}, M.~R. and {Levan}, A.~J.},
        title = "{(No) dynamical constraints on the mass of the black hole in two ULXs}",
      journal = {Astronomische Nachrichten},
         year = 2011,
        month = may,
       volume = {332},
       number = {4},
        pages = {398},
          doi = {10.1002/asna.201011508},
archivePrefix = {arXiv},
       eprint = {1011.2155},
 primaryClass = {astro-ph.HE},
       adsurl = {https://ui.adsabs.harvard.edu/abs/2011AN....332..398R}
}

@ARTICLE{zhou2022,
       author = {{Zhou}, Changxing and {Bian}, Fuyan and {Feng}, Hua and {Huang}, Jiahui},
        title = "{Very Large Telescope MUSE Observations of the Bubble Nebula around NGC 1313 X-2 and Evidence for Additional Photoionization}",
      journal = {\apj},
         year = 2022,
        month = aug,
       volume = {935},
       number = {1},
          eid = {38},
        pages = {38},
          doi = {10.3847/1538-4357/ac815f},
archivePrefix = {arXiv},
       eprint = {2207.06577},
 primaryClass = {astro-ph.HE},
       adsurl = {https://ui.adsabs.harvard.edu/abs/2022ApJ...935...38Z}
}

@INPROCEEDINGS{castellikurucz03,
       author = {{Castelli}, F. and {Kurucz}, R.~L.},
        title = "{New Grids of ATLAS9 Model Atmospheres}",
    booktitle = {Modelling of Stellar Atmospheres},
         year = 2003,
       editor = {{Piskunov}, N. and {Weiss}, W.~W. and {Gray}, D.~F.},
       series = {IAU Symposium},
       volume = {210},
        month = jan,
        pages = {A20},
          doi = {10.48550/arXiv.astro-ph/0405087},
archivePrefix = {arXiv},
       eprint = {astro-ph/0405087},
 primaryClass = {astro-ph},
       adsurl = {https://ui.adsabs.harvard.edu/abs/2003IAUS..210P.A20C}
}

@ARTICLE{ripamonti11,
       author = {{Ripamonti}, E. and {Mapelli}, M. and {Zampieri}, L. and {Colpi}, M.},
        title = "{The metallicity of the nebula surrounding the ultra-luminous X-ray source NGC 1313 X-2}",
      journal = {Astronomische Nachrichten},
         year = 2011,
        month = may,
       volume = {332},
       number = {4},
        pages = {418-421},
          doi = {10.1002/asna.201011512},
archivePrefix = {arXiv},
       eprint = {1009.5708},
 primaryClass = {astro-ph.CO},
       adsurl = {https://ui.adsabs.harvard.edu/abs/2011AN....332..418R}
}

@ARTICLE{gies02,
       author = {{Gies}, D.~R. and {Huang}, W. and {McSwain}, M.~V.},
        title = "{The Spectrum of the Mass Donor Star in SS 433}",
      journal = {\apjl},
         year = 2002,
        month = oct,
       volume = {578},
       number = {1},
        pages = {L67-L70},
          doi = {10.1086/344436},
archivePrefix = {arXiv},
       eprint = {astro-ph/0208044},
 primaryClass = {astro-ph},
       adsurl = {https://ui.adsabs.harvard.edu/abs/2002ApJ...578L..67G}
}

@ARTICLE{hillwig04,
       author = {{Hillwig}, T.~C. and {Gies}, D.~R. and {Huang}, W. and {McSwain}, M.~V. and {Stark}, M.~A. and {van der Meer}, A. and {Kaper}, L.},
        title = "{Identification of the Mass Donor Star's Spectrum in SS 433}",
      journal = {\apj},
         year = 2004,
        month = nov,
       volume = {615},
       number = {1},
        pages = {422-431},
          doi = {10.1086/423927},
archivePrefix = {arXiv},
       eprint = {astro-ph/0403634},
 primaryClass = {astro-ph},
       adsurl = {https://ui.adsabs.harvard.edu/abs/2004ApJ...615..422H}
}

@ARTICLE{eggleton1983,
       author = {{Eggleton}, P.~P.},
        title = "{Aproximations to the radii of Roche lobes.}",
      journal = {\apj},
         year = 1983,
        month = may,
       volume = {268},
        pages = {368-369},
          doi = {10.1086/160960},
       adsurl = {https://ui.adsabs.harvard.edu/abs/1983ApJ...268..368E}
}

@ARTICLE{sepinsky07,
       author = {{Sepinsky}, J.~F. and {Willems}, B. and {Kalogera}, V.},
        title = "{Equipotential Surfaces and Lagrangian Points in Nonsynchronous, Eccentric Binary and Planetary Systems}",
      journal = {\apj},
         year = 2007,
        month = may,
       volume = {660},
       number = {2},
        pages = {1624-1635},
          doi = {10.1086/513736},
archivePrefix = {arXiv},
       eprint = {astro-ph/0612508},
 primaryClass = {astro-ph},
       adsurl = {https://ui.adsabs.harvard.edu/abs/2007ApJ...660.1624S}
}

@ARTICLE{qiufeng2021,
       author = {{Qiu}, Yanli and {Feng}, Hua},
        title = "{Constraining Soft and Hard X-Ray Irradiation in Ultraluminous X-Ray Sources}",
      journal = {\apj},
         year = 2021,
        month = dec,
       volume = {922},
       number = {2},
          eid = {91},
        pages = {91},
          doi = {10.3847/1538-4357/ac3569},
archivePrefix = {arXiv},
       eprint = {2111.02237},
 primaryClass = {astro-ph.HE},
       adsurl = {https://ui.adsabs.harvard.edu/abs/2021ApJ...922...91Q}
}

@ARTICLE{seheon2018,
       author = {{Oh}, Se-Heon and {Staveley-Smith}, Lister and {Spekkens}, Kristine and {Kamphuis}, Peter and {Koribalski}, B{\"a}rbel S.},
        title = "{2D Bayesian automated tilted-ring fitting of disc galaxies in large H I galaxy surveys: 2DBAT}",
      journal = {\mnras},
         year = 2018,
        month = jan,
       volume = {473},
       number = {3},
        pages = {3256-3298},
          doi = {10.1093/mnras/stx2304},
archivePrefix = {arXiv},
       eprint = {1709.02049},
 primaryClass = {astro-ph.GA},
       adsurl = {https://ui.adsabs.harvard.edu/abs/2018MNRAS.473.3256O}
}

@ARTICLE{zampieri2012,
       author = {{Zampieri}, L. and {Impiombato}, D. and {Falomo}, R. and {Gris{\'e}}, F. and {Soria}, R.},
        title = "{Searching for the orbital period of the ultraluminous X-ray source NGC 1313 X-2}",
      journal = {\mnras},
         year = 2012,
        month = jan,
       volume = {419},
       number = {2},
        pages = {1331-1337},
          doi = {10.1111/j.1365-2966.2011.19790.x},
archivePrefix = {arXiv},
       eprint = {1109.4437},
 primaryClass = {astro-ph.HE},
       adsurl = {https://ui.adsabs.harvard.edu/abs/2012MNRAS.419.1331Z}
}

@ARTICLE{evans2007,
       author = {{Evans}, P.~A. and {Beardmore}, A.~P. and {Page}, K.~L. and {Tyler}, L.~G. and {Osborne}, J.~P. and {Goad}, M.~R. and {O'Brien}, P.~T. and {Vetere}, L. and {Racusin}, J. and {Morris}, D. and {Burrows}, D.~N. and {Capalbi}, M. and {Perri}, M. and {Gehrels}, N. and {Romano}, P.},
        title = "{An online repository of Swift/XRT light curves of {\ensuremath{\gamma}}-ray bursts}",
      journal = {\aap},
         year = 2007,
        month = jul,
       volume = {469},
       number = {1},
        pages = {379-385},
          doi = {10.1051/0004-6361:20077530},
archivePrefix = {arXiv},
       eprint = {0704.0128},
 primaryClass = {astro-ph},
       adsurl = {https://ui.adsabs.harvard.edu/abs/2007A&A...469..379E}
}

@INPROCEEDINGS{cardelli1989,
       author = {{Cardelli}, J.~A. and {Clayton}, G.~C. and {Mathis}, J.~S.},
        title = "{The relationship between IR, optical, and UV extinction.}",
    booktitle = {Interstellar Dust},
         year = 1989,
       editor = {{Allamandola}, Louis J. and {Tielens}, A.~G.~G.~M.},
       series = {IAU Symposium},
       volume = {135},
        month = dec,
        pages = {5-10},
       adsurl = {https://ui.adsabs.harvard.edu/abs/1989IAUS..135P...5C}
}

@ARTICLE{moorwood_1996,
       author = {{Moorwood}, Alan F.~M.},
        title = "{Starburst Galaxies}",
      journal = {\ssr},
         year = 1996,
        month = aug,
       volume = {77},
       number = {3-4},
        pages = {303-366},
          doi = {10.1007/BF00226226},
       adsurl = {https://ui.adsabs.harvard.edu/abs/1996SSRv...77..303M}
}

@ARTICLE{weaver1977,
       author = {{Weaver}, R. and {McCray}, R. and {Castor}, J. and {Shapiro}, P. and {Moore}, R.},
        title = "{Interstellar bubbles. II. Structure and evolution.}",
      journal = {\apj},
         year = 1977,
        month = dec,
       volume = {218},
        pages = {377-395},
          doi = {10.1086/155692},
       adsurl = {https://ui.adsabs.harvard.edu/abs/1977ApJ...218..377W}}

@ARTICLE{uttley_2014,
       author = {{Uttley}, P. and {Cackett}, E.~M. and {Fabian}, A.~C. and {Kara}, E. and {Wilkins}, D.~R.},
        title = "{X-ray reverberation around accreting black holes}",
      journal = {\aapr},
         year = 2014,
        month = aug,
       volume = {22},
          eid = {72},
        pages = {72},
          doi = {10.1007/s00159-014-0072-0},
archivePrefix = {arXiv},
       eprint = {1405.6575},
 primaryClass = {astro-ph.HE},
       adsurl = {https://ui.adsabs.harvard.edu/abs/2014A&ARv..22...72U}
}
\bsp	
\label{lastpage}

\end{document}